\documentclass[11pt,a4paper]{article}
\pdfoutput=1
\usepackage{graphicx}
\usepackage{amsfonts}
\usepackage[margin=2cm]{geometry}
\usepackage{amsmath,amssymb,amsthm,array,booktabs,cite,dsfont,mathtools,multirow,placeins,rotating,stmaryrd,tensor,verbatim,subfig}
\usepackage[bookmarksnumbered,linktocpage,pdfstartview=FitH]{hyperref}
\usepackage{dcolumn}
\usepackage[all]{hypcap}

\usepackage{bm}
\usepackage{cancel}
\usepackage{amscd,slashed}

\newcolumntype{M}[1]{>{$}{#1}<{$}}

\newcommand{\sst}[1]{{\scriptscriptstyle #1}}

\newcommand{\rep}[1]{\ensuremath{\mathbf{#1}}}

\def\0{{\sst{(0)}}}
\def\1{{\sst{(1)}}}
\def\2{{\sst{(2)}}}
\def\3{{\sst{(3)}}}
\def\4{{\sst{(4)}}}
\def\5{{\sst{(5)}}}
\def\6{{\sst{(6)}}}
\def\7{{\sst{(7)}}}

\newcommand{\be}{\begin{equation}}
\newcommand{\ee}{\end{equation}}
\def\ba{\begin{array}}
\def\ea{\end{array}}

\newcommand{\bea}{\begin{eqnarray}}
\newcommand{\eea}{\end{eqnarray}}

\newcommand{\ox}{\otimes}
\newcommand{\eps}{\varepsilon}

\DeclareMathOperator{\tr}{tr}

\DeclareMathOperator{\Hom}{Hom}
\DeclareMathOperator{\SO}{SO}
\DeclareMathOperator{\Orth}{O}
\DeclareMathOperator{\USp}{USp}
\DeclareMathOperator{\SL}{SL}
\DeclareMathOperator{\SU}{SU}
\DeclareMathOperator{\Sp}{Sp}

\DeclareMathOperator{\Un}{U}

\newcommand{\susy}{\mathcal{N}}

\newcommand{\blf}[2]{\langle#1 , #2\rangle}
\newcommand{\bn}{\mathbf{n}}

\newcommand{\tri}{\mathfrak{tri}}
\newcommand{\alg}{\mathds{A}}
\newcommand{\al}{\mathds{A}}

\newcommand{\Al}{\mathds{A}}
\newcommand{\mf}{\mathfrak}
\newcommand{\F}{\mathds{F}}
\newcommand{\R}{\mathds{R}}
\newcommand{\C}{\mathds{C}}
\newcommand{\Q}{\mathds{H}}
\newcommand{\Oct}{\mathds{O}}
\newcommand{\Z}{\mathds{Z}}
\newcommand{\sigmabar}{\bar{\sigma}}
\newcommand{\RCHO}{\mathds{R},\mathds{C},\mathds{H},\mathds{O}}
\newcommand{\id}{\mathds{1}}
\newcommand{\N}{\mathcal{N}}

\begin{document}

\begin{titlepage}
\begin{center}
\hfill Imperial/TP/2013/mjd/03\\

\vskip 2cm

{\Huge \bf A magic pyramid of supergravities}

\vskip 1.5cm

{\bf A. Anastasiou, L.~Borsten, M.~J.~Duff, L.~J.~Hughes and
S.~Nagy}

\vskip 20pt

{\it Theoretical Physics, Blackett Laboratory, Imperial College London,\\
 London SW7 2AZ, United Kingdom}\\\vskip 5pt
\texttt{alexandros.anastasiou07@imperial.ac.uk}\\
\texttt{leron.borsten@imperial.ac.uk}\\
\texttt{m.duff@imperial.ac.uk}\\
\texttt{leo.hughes07@imperial.ac.uk}\\
\texttt{s.nagy11@imperial.ac.uk}

\end{center}

\vskip 2.2cm

\begin{center} {\bf ABSTRACT}\\[3ex]\end{center}

By formulating $\mathcal{N}=1,2,4, 8$, $D=3$,  Yang-Mills  with a single Lagrangian and single set of transformation rules, but with fields valued respectively in $\R,\C,\Q, \Oct$,  it was recently shown that  tensoring left and right multiplets yields a Freudenthal-Rosenfeld-Tits magic square of $D=3$ supergravities. This was subsequently tied in  with the more familiar $\R,\C,\Q, \Oct$ description of spacetime to give a unified division-algebraic description of extended super Yang-Mills in $D=3,4,6,10$. Here, these constructions are brought together resulting in a \emph{magic pyramid} of supergravities. The base of the  pyramid in $D=3$ is the known $4 \times4$ magic square,  while the higher levels are comprised of a $3 \times 3$ square in $D=4$, a $2 \times 2$ square in $D=6$ and Type II supergravity at the apex in $D=10$.
 The corresponding U-duality groups are given by a new algebraic structure, the magic pyramid formula, which may be regarded as being defined over three division algebras, one for spacetime and  each of the left/right Yang-Mills multiplets. We also construct a \emph{conformal magic pyramid} by tensoring conformal supermultiplets in $D=3,4,6$. The missing entry in $D=10$ is suggestive of an exotic theory with $G/H$ duality structure $F_{4(4)}/\Sp(3) \times \Sp(1)$.



\vfill


\end{titlepage}

\newpage \setcounter{page}{1} \numberwithin{equation}{section} \tableofcontents

\section{Introduction}

In recent years  gauge and gravitational scattering amplitudes have undergone something of a renaissance \cite{Elvang:2013cua}, resulting not only in dramatic computational advances but also important conceptual insights. One such development,  straddling both the technical and conceptual,  is the colour-kinematic duality of gauge amplitudes introduced by Bern, Carrasco and Johansson  \cite{Bern:2008qj}. Exploiting this duality it has been shown that gravitational amplitudes may be reconstructed using a double-copy of gauge amplitudes  suggesting a possible interpretation of perturbative gravity as ``the square of Yang-Mills'' \cite{Bern:2010ue, Bern:2010yg}. This perspective has proven itself remarkably effective, rendering possible previously intractable gravitational scattering amplitude calculations   \cite{Bern:2009kd}; it is both   conceptually suggestive and technically advantageous. Yet, the idea of gravity as the square of Yang-Mills  is not specific to amplitudes, having appeared previously in a number of different,  but sometimes related, contexts \cite{Kawai:1985xq, Green:1987sp, Sen:1995ff, Antoniadis:1992sa, Carrasco:2012ca}. While it would seem  there is now a growing web of relations connecting gravity to  ``gauge $\times$ gauge", it is as yet not   clear to what  extent gravity may be regarded as the  square of Yang-Mills.

Here, we ask how the non-compact global symmetries of supergravity \cite{Cremmer:1979up}, or in an M-theory context the so-called U-dualities \cite{Duff:1990hn, Hull:1994ys}, might be related to the ``square'' of those in super Yang-Mills (SYM), namely R-symmetries. Surprisingly,  in the course of addressing this question the division algebras $\alg=\R, \C, \Q, \Oct$ and their associated symmetries  reveal themselves as playing an intriguing role. Tensoring, as in \cite{deWit:2002vz}, $\susy_L$ and $\susy_R$ super Yang-Mills multiplets in $D=3,4,6,10$ dimensions yields supergravities with U-dualities given by a \emph{magic pyramid formula} parametrized by a triple of division algebras $(\alg_{n}, \alg_{n \mathcal{N}_L}, \alg_{n\mathcal{N}_R})$, one for spacetime and two for the left/right Yang-Mills multiplets.

In  previous work \cite{Borsten:2013bp} we built a symmetric $4\times 4$ array of   three-dimensional supergravity multiplets, with $\mathcal{N}=\mathcal{N}_L+\mathcal{N}_R$, by tensoring a left $\mathcal{N}_L=1,2,4,8$ SYM multiplet with a right $\mathcal{N}_R=1,2,4,8$ SYM multiplet. Remarkably, the corresponding U-dualities filled out the Freudenthal-Rosenfeld-Tits magic square \cite{Freudenthal:1954, Freudenthal:1959, Freudenthal:1964, Rosenfeld:1956,Tits:1966,Vinberg:1966};  a symmetric $4\times 4$ array of Lie algebras  defined by a single formula taking as its argument a pair of division algebras, 
\be\label{eq:tri}
\mf{L}_3(\alg_{ \mathcal{N}_L}, \alg_{ \mathcal{N}_R}):=\mf{tri}(\alg_{ \mathcal{N}_L})\oplus \mf{tri}(\alg_{ \mathcal{N}_R})+3(\alg_{ \mathcal{N}_L}\otimes\alg_{ \mathcal{N}_R}),
\ee
where the subscripts denote the dimension of the algebras. See  \autoref{tab:ms1}. Here, $\mf{tri}(\alg)$ denotes the triality Lie algebra of $\alg$, a generalisation of the algebra of derivations which contains as a sub-algebra the R-symmetry  of super Yang-Mills.  See \autoref{sec:ms}.
 
  \begin{table}[ht]
 \begin{center}
\begin{tabular}{c|ccccccc}
\hline
\hline
 $\alg_L/\alg_R$ && $\R$ & $\C$  & $\Q$  & $\Oct$ & \\
 \hline
 \\
   $\R$ && $\SL(2, \R)$ & $\SU(2,1)$   & $\USp(4,2)$   & $F_{4(-20)}$  & \\
  $\C$ && $\SU(2,1)$ & $\SU(2,1)\times  \SU(2,1)$   & $\SU(4,2)$   & $E_{6(-14)}$   &\\
  $\Q$ && $\USp(4,2)$ & $\SU(4,2)$   & $\SO(8,4)$   & $E_{7(-5)}$  & \\
   $\Oct$ && $F_{4(-20)}$ & $E_{6(-14)}$   & $E_{7(-5)}$   & $E_{8(8)}$&   \\
   \\
   \hline
   \hline
\end{tabular}
\caption[Magic square of required real forms.]{The magic square \label{tab:ms1}}
 \end{center}
\end{table}

\newpage
The Freudenthal-Rosenfeld-Tits magic square\footnote{There are a number of equivalent forms/constructions of the magic square formula \eqref{eq:tri} due to, amongst others, Tits \cite{Tits:1966}, Vinberg \cite{Vinberg:1966}, Kantor \cite{kantor1973models} and Barton-Sudbery \cite{Barton:2003}. The ternary algebra approach of \cite{kantor1973models} was generalised by Bars-G{\"u}naydin \cite{Bars:1978yx} to include super Lie algebras.   The form given in  \eqref{eq:tri} is due to Barton-Sudbery. We adopt this construction, modified as a Lie algebra to produce the required real forms presented in \autoref{tab:ms1}, as will be explained in \autoref{sec:ms}. This specific square of real forms was first derived in \cite{Cacciatori:2012cb} using Tits' formula defined over a Lorentzian Jordan algebra. By, for example, altering the signature of the algebras  a variety of real forms  can be accommodated. See \cite{Cacciatori:2012cb}  for a comprehensive account in the  context of supergravity.}  historically originated  from efforts to understand the exceptional  Lie groups in terms of octonionic geometries and, accordingly,  the scalar fields of the corresponding supergravities  parametrize division algebraic projective spaces
 \cite{Borsten:2013bp}. The connection to the division algebras in fact goes deeper; the appearance of the magic square can be explained using the observation that the $D=3$, $\mathcal{N}=1,2,4, 8$ Yang-Mills theories can be formulated with a single Lagrangian and a single set of transformation rules, using fields  valued in $\R,\C,\Q$ and $\Oct$, respectively. Tensoring an $\mathds{A}_{\mathcal{N}_L}$-valued super Yang-Mills multiplet with an $\mathds{A}_{\mathcal{N}_L}$-valued super Yang-Mills multiplet yields a supergravity mulitplet with fields valued in $\mathds{A}_{\mathcal{N}_L}\otimes \mathds{A}_{\mathcal{N}_L}$, making a magic square of U-dualities appear rather natural.

Of course, the connection between supersymmetry and  division algebras is not new.  In particular, Kugo and Townsend \cite{Kugo:1982bn} related the existence of minimal $\mathcal{N}=1$ super Yang-Mills multiplets in only three, four, six and ten dimensions  directly  to the existence of only four division algebras $\R,\C,\Q$ and  $\Oct$, an observation which  has been subsequently developed in a variety of directions. See, for example, \cite{Sudbery:1984,Evans:1987tm,Duff:1987qa,Blencowe:1988sk, Berkovits:1993hx, Schray:1994ur, Manogue:1998rv, Baez:2001dm,Toppan:2003yx, Gunaydin:2005zz, Kuznetsova:2006ws, Borsten:2008wd, Baez:2009xt, Rios:2011fa, Anastasiou:2013cya} and the references therein. From this point of view the division algebras are related to the spacetime symmetries, rather than the internal R-symmetries, via the Lie algebra isomorphism (in the sense of \cite{Sudbery:1984})
 \be\mathfrak{sl}(2, \alg_n)\cong \mathfrak{so}(1, n+2).\ee 
 Indeed, the unique $D=10$, $\mathcal{N}=1$ super Yang-Mills theory can be formulated using octonionic spacetime fields \cite{Baez:2009xt}. By dimensionally reducing this octonionic theory, which corresponds to Cayley-Dickson halving,  one recovers the octonionic formulation of $D=3$, $\mathcal{N}=8$ Yang-Mills presented in \cite{Borsten:2013bp}, tying together the division algebraic descriptions of spacetime and supersymmetry. This approach gives a unified division algebraic description of ($D=3$, $\mathcal{N}=1,2,4,8$), ($D=4$, $\mathcal{N}=1,2,4$), ($D=6$, $\mathcal{N}=1,2$) and ($D=10$, $\mathcal{N}=1$)  Yang-Mills theories. A given $(D=n+2$, $\mathcal{N})$ theory (the field content, Lagrangian and transformation rules) is completely specified  by selecting an ordered pair $\alg_n\subseteq \alg_{n\susy}$, where again the subscripts denote the dimension of the algebras \cite{Anastasiou:2013cya}. This unity  is neatly expressed through the fact that the (modified) triality algebras appearing in \eqref{eq:tri} are the direct sum of the spacetime little group and the internal R-symmetry algebras \cite{Anastasiou:2013cya}. 

\begin{figure}\label{fig:p1} 
\begin{center}
\includegraphics[scale=0.12]{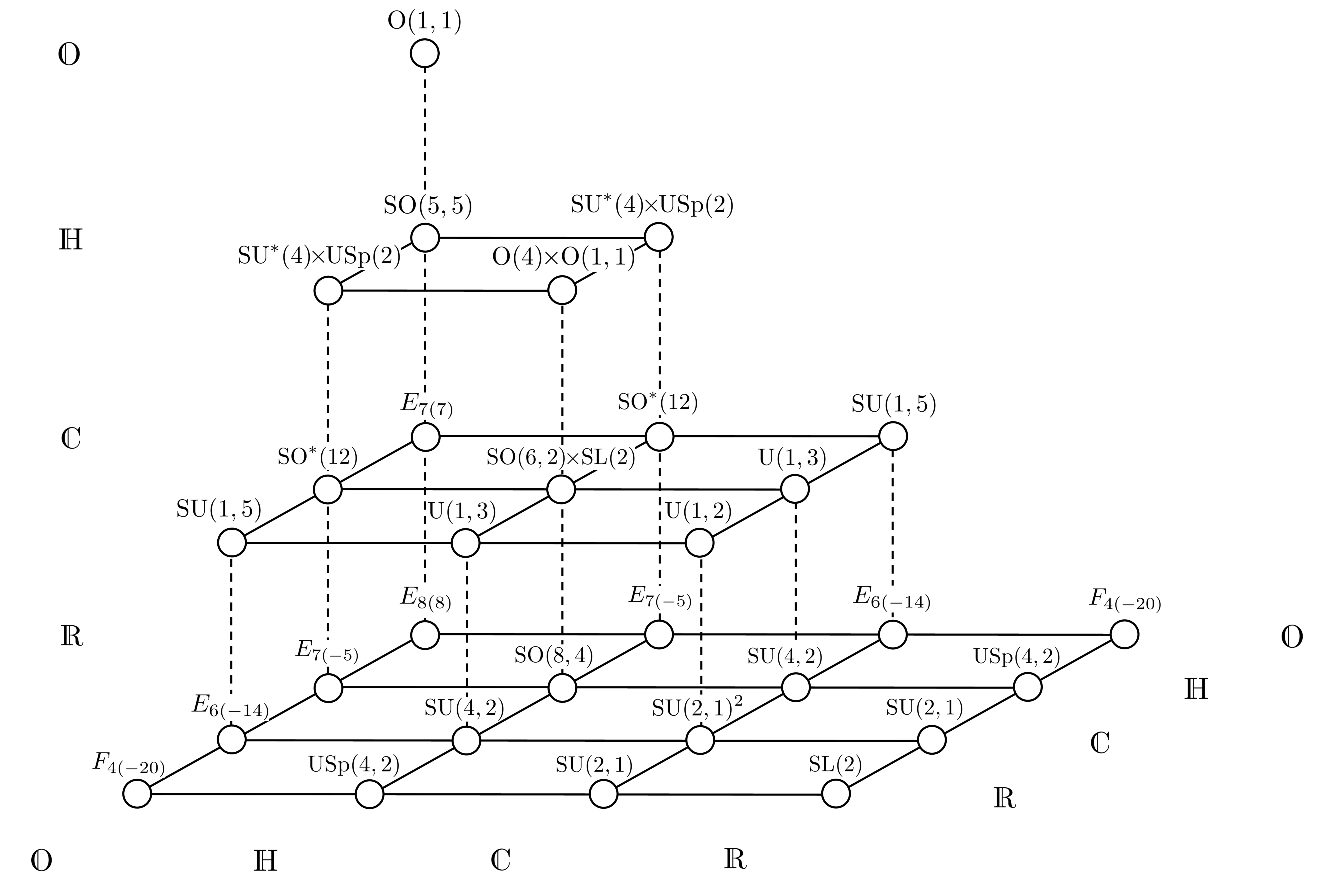}
\caption{\footnotesize{A magic pyramid of supergravities. The vertical axis labels the spacetime division algebra $\Al_n$, while the horizontal axes label the algebras associated with the number of supersymmetries $\alg_{n \mathcal{N}_L}$ and $\alg_{n\mathcal{N}_R}$. }}\label{pyramid}
\end{center}
\end{figure}

\begin{figure}
\begin{center}
\includegraphics[scale=0.11]{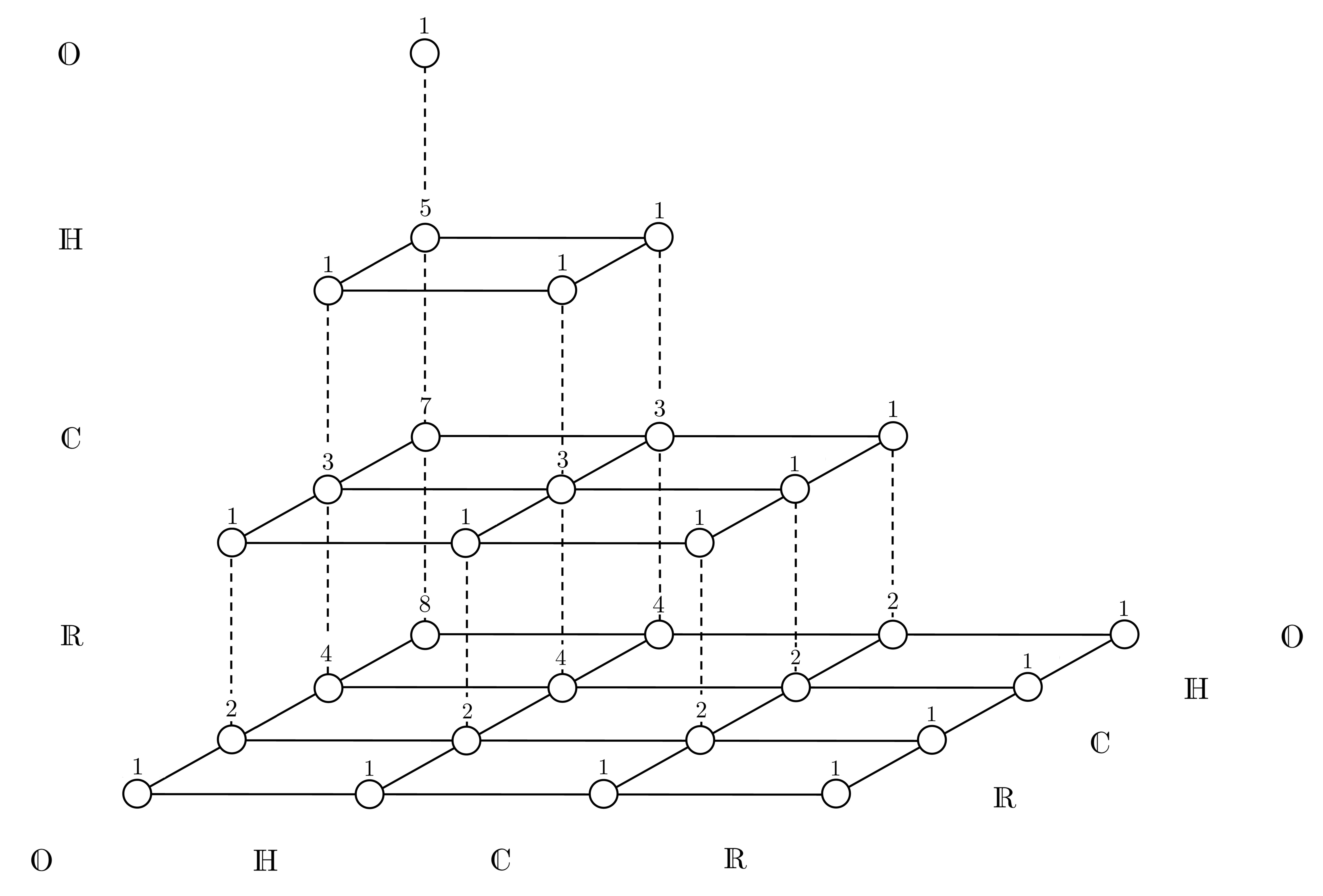}
\caption{\footnotesize{The ranks of the scalar cosets $G/H$, where $G$ is the U-duality and $H$ is its maximal compact subgroup (the entries here apply to the original magic pyramid obtained by squaring SYM rather that from squaring conformal theories). }}\label{cosets}
\end{center}
\end{figure}

In the present work we bring together the various roles of the division algebras discussed above to construct a pyramid of supergravities by tensoring left/right $\alg_{n\susy_L}/\alg_{n\susy_R}$-valued Yang-Mills multiplets in $D=n+2$. The base of the pyramid in $D=3$ is the $4\times 4$ magic square of supergravities,   with a $3 \times 3$ square in $D=4$, a $2 \times 2$ square in $D=6$ and Type II supergravity at the apex in $D=10$. The totality defines a new algebraic structure: the magic pyramid. The U-dualities are given by the  magic pyramid formula, 
\be\label{eq:pyramidformula}
\mathfrak{Pyr}(\alg_{n}, \alg_{n\susy_L}, \alg_{n\susy_R}):=\left\{u\in\mf{L}_3(\alg_{n\N_L}, \alg_{n\N_R})-\mf{so}(\alg_n)_{ST}\Big|[u,\mf{so}(\alg_n)_{ST}]=0\right\},
\ee
where $\mf{so}(\alg_n)_{ST}\subset\mf{L}_3(\alg_{n\N_L}, \alg_{n\N_R})$ is the subalgebra of on-shell spacetime transformations (the spacetime little group\footnote{We neglect the translation generators of ISO$(D-2)$ since they annihilate  physical states. Note, throughout we do not distinguish the special orthogonal group from its double cover (SO vs. Spin) for typographical clarity. Of course, this is an important distinction and we hope that this  rather non-trivial abuse of notation will not cause confusion given the context.}  in $D=n+2$ dimensions is $\SO(n)\cong\SO(\alg_n)$,  as described in \autoref{sec:ms}). This is the natural generalisation of the magic square formula given in \eqref{eq:tri}; the largest subalgebra of $\mf{L}_3(\alg_{n \mathcal{N}_L}, \alg_{n \mathcal{N}_R})$ that respects spacetime transformations generates the U-duality. See \autoref{pyramid} for the  magic pyramid of U-duality groups described by \eqref{eq:pyramidformula}, and \autoref{cosets} for the ranks of the corresponding cosets, which we include to highlight the  the curious pattern they follow.

The pyramid formula may also be understood geometrically. As observed in \cite{deWit:1992up} for the exceptional cases, the $D=3$ Freudenthal magic square can  be  regarded  as the isometries of  the division algebraic projective spaces $(\alg_{\mathcal{N}_L}\otimes\alg_{ \mathcal{N}_R})\mathds{P}^2$. Here we are being rather heuristic - for  more detailed and elegant  treatments of magic square projective geometry see \cite{Besse:1987, Baez:2001dm, Landsberg:2001, atiyah2002projective} and the references therein. In essence the pyramid algebra describes the isometries of special submanifolds of these projective spaces. On tensoring   SYM multiplets in $D>3$, we must identify a diagonal $\alg_n$ subalgebra to be associated with spacetime. This can be thought of as introducing an $\alg_{n}$-structure on the $\SO(\mathcal{N}_L +{\mathcal{N}_R})$ that acts (in the spinor representation) on the tangent space $(\alg_{\N_L}\otimes\alg_{\N_R})^2$ at each point on the projective plane. This splits the projective space into two pieces, one internal and one spacetime. The isometries of the internal component yield the magic pyramid, while the remaining symmetries generate the spacetime little group. 

\begin{figure}[t]
\begin{center}
\includegraphics[scale=0.12]{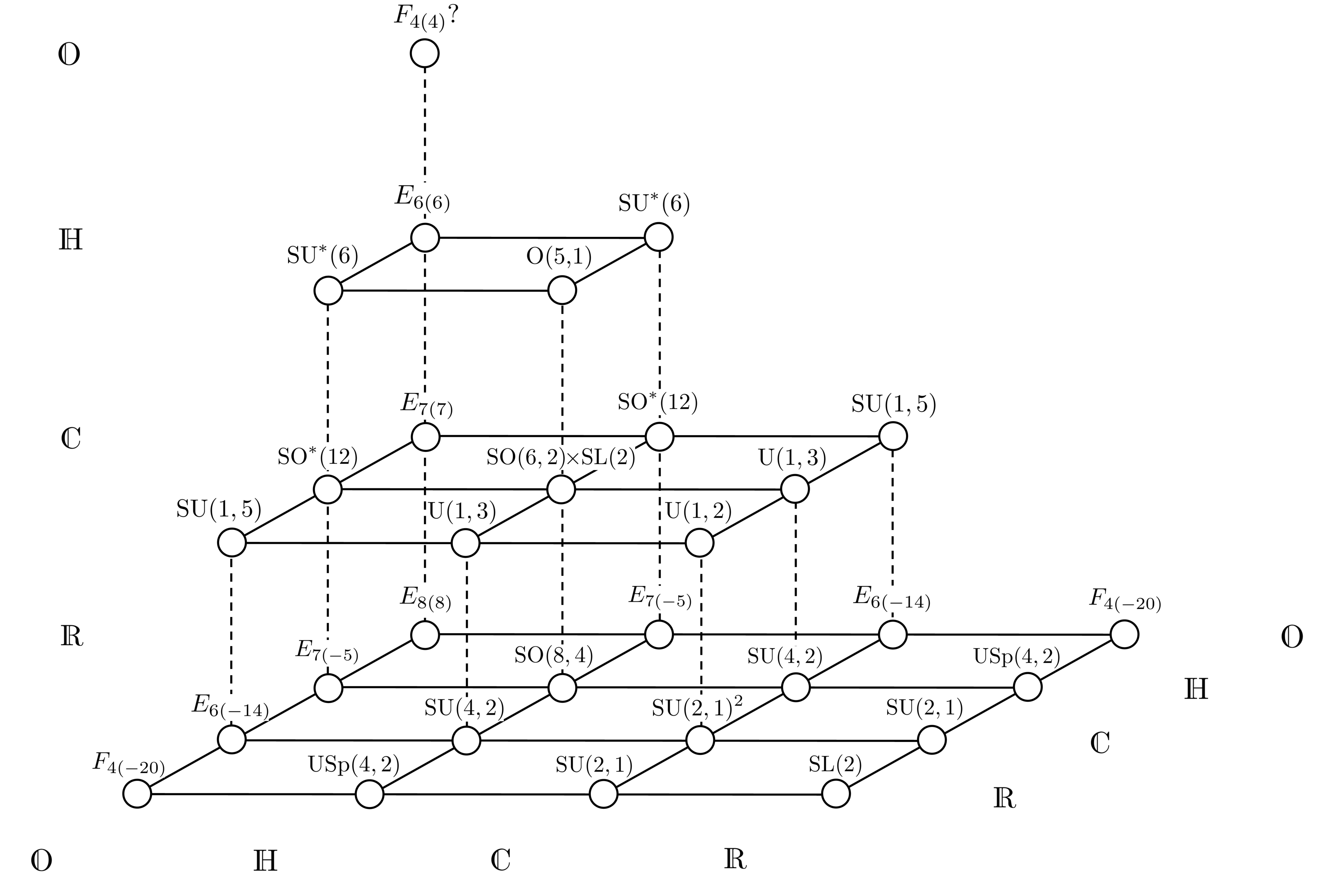}
\caption{\footnotesize{The conformal magic pyramid. Note, the exterior faces, up to real forms, are given by the magic square (i.e. the $4\times 4$ base) cut across its diagonal.}}\label{fig:confp1} 
\end{center}
\end{figure}

Rather than   SYM   one might also consider ``squaring'' the $D=3,4,6$ conformal multiplets: super Chern-Simons-matter (CSm), SYM and tensor, respectively \cite{deWit:2002vz}. This  yields another magic pyramid, as described in \autoref{sec:conf}, which we will refer to as the conformal pyramid. See  \autoref{fig:confp1}. It has the remarkable property that its faces are also given by the known $D=3$ magic square. For example, trading  the maximal super Yang-Mills  in $D=6$ for the $(2,0)$ tensor mulitplet swaps the resulting maximal supergravity with $\SO(5,5)$ U-duality for the  non-gravitational $(4,0)$  self-dual-Weyl multiplet with $E_{6(6)}$ U-duality considered in \cite{Hull:2000zn, deWit:2002vz, Chiodaroli:2011pp}.  Given the recent progress in understanding three-dimensional supergravity amplitudes  as double copies of Bagger-Lambert-Gustavsson theories \cite{Huang:2012wr}, one might anticipate applications  to this line of enquiry. More speculatively, the conformal pyramid in $D=3,4,6$ suggests an exotic $D=10$ theory with global symmetry $F_{4(4)}$, although it would have to be highly non-conventional (even heretical) from the standard perspective on the classification of  supermultiplets. For earlier appearances of $F_4$ in 10 and 11 dimensions see \cite{Pengpan:1998qn, Ramond:2001ud, Ramond:2000ea, Duff:2002rw, Sati:2008fz}.

Note, in the present  paper  the complete Freudenthal-Rosenfeld-Tits magic square describes the U-dualities of  conventional  $D=3$ supergravities.  Its role here is not to be confused with its appearance in the  important, and aptly named, ``magic supergravities'' of G{\"u}naydin-Sierre-Townsend \cite{Gunaydin:1983bi, Gunaydin:1983rk}. In this context the $\C, \Q,$ and $\Oct$ rows  of the magic square (with a different set of real forms) describe the U-dualities of the magic supergravities  in $D=5,4$ and $3$ respectively.  The magic square   also appeared previously  in a further, distinct, supersymmetric setting  in \cite{Gunaydin:1975mp}.

In \autoref{sec:ms} we review division algebras and the square construction, giving the details of our formulation of \autoref{tab:ms1}, which were omitted from \cite{Borsten:2013bp}. In \autoref{sec:mp} we briefly recall the division algebraic description of  SYM and then construct the magic pyramid of supergravitites. In \autoref{sec:conf} we introduce the conformal pyramid.

\section{The magic square}\label{sec:ms}

An algebra $\mathds{A}$ defined over  $\R$ with identity element $e_0$, is said to be \emph{composition} if it has a non-degenerate quadratic form\footnote{A \emph{quadratic norm} on a vector space $V$ over a field $\R$ is a map $\bn:V\to\R$ such that: (1) $\bn(\lambda a)=\lambda^2\bn(a),  \lambda\in\R, a\in V$ and 
(2)
$
\langle a, b\rangle:=\bn(a+b)-\bn(a)-\bn(b)
$
is bilinear.}
$\bn:\mathds{A}\to\R$ such that,
\begin{equation}
\bn(ab)=\bn(a)\bn(b),\quad \forall~~ a,b \in\alg,
\end{equation}
where we denote the multiplicative product of the algebra by juxtaposition.

A composition algebra $\mathds{A}$  is said to be \emph{division} if it contains no zero divisors,
\begin{equation*}
ab=0\quad \Rightarrow\quad a=0\quad\text{or}\quad b=0,
\end{equation*}
in which case $\bn$ is positive semi-definite and $\alg$ is referred to as a normed division algebra.
Hurwitz's celebrated theorem states that there are exactly four normed division algebras \cite{Hurwitz:1898}: the reals, complexes, quaternions and octonions, denoted respectively by $\R, \C, \Q$ and $\Oct$. 

Regarding ${\R}\subset\alg$ as the scalar multiples of the identity $ \R e_0$ we may decompose $\mathds{A}$ into its ``real'' and ``imaginary'' parts $\alg={\R}\oplus \alg'$, where $\alg'\subset\alg$ is the subspace orthogonal to $\R$. An arbitrary element $a\in\alg$ may be written $a=\text{Re}(a) +\text{Im}(a)$. Here  $\text{Re}(a)\in\R e_0$, $\text{Im}(a)\in \alg'$ and
\be
\text{Re}(a)=\frac{1}{2}(a+\overline{a}), \qquad \text{Im}(a)=\frac{1}{2}(a-\overline{a}),
\ee
where we have defined conjugation  using the bilinear form,
\be
\overline{a}:=\blf{a}{e_0}e_0-a, \qquad \langle a, b\rangle:=\bn(a+b)-\bn(a)-\bn(b).
\ee 

\begin{figure}[h!]
  \centering
    \includegraphics[width=0.4\textwidth]{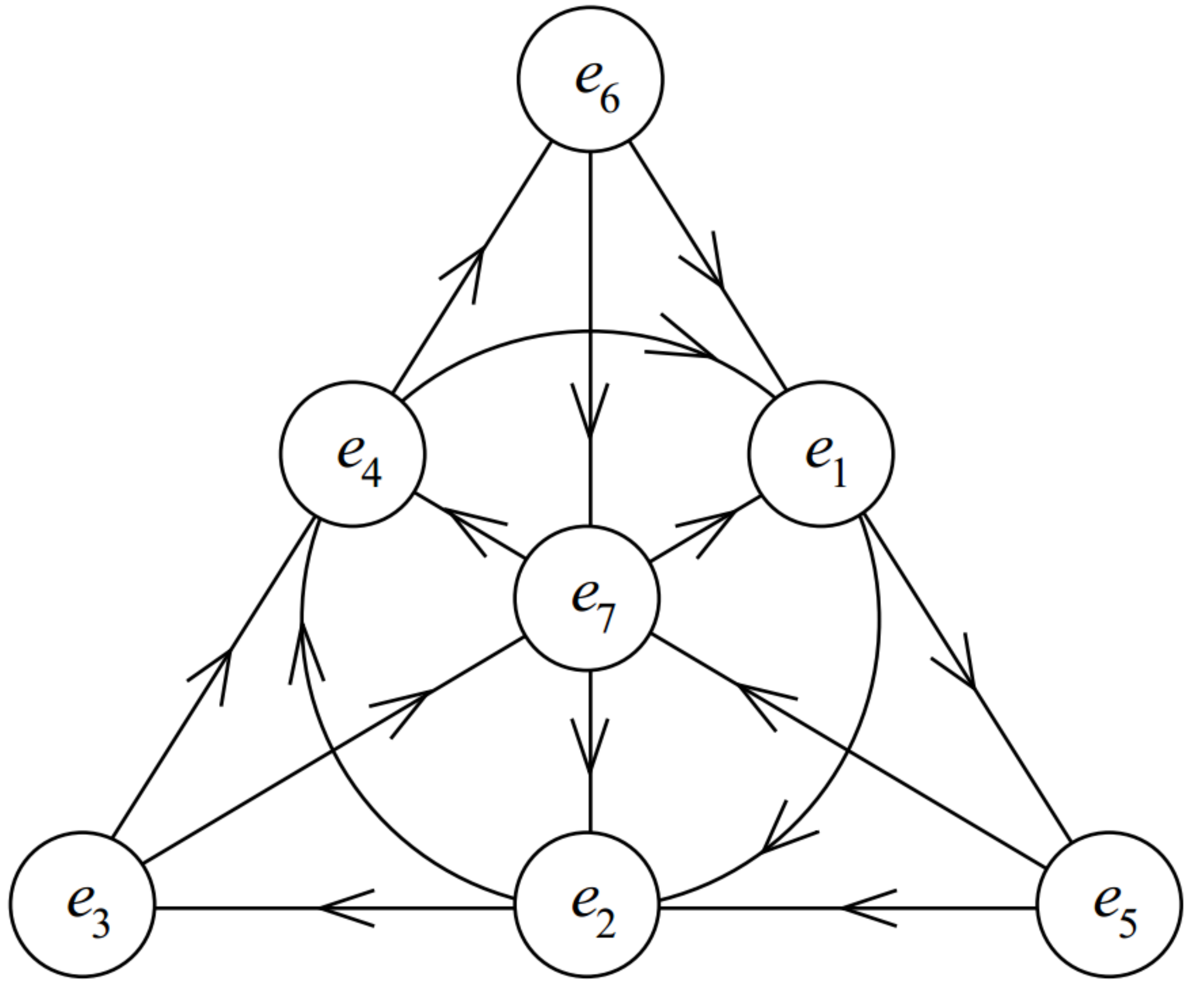}
  \caption{\footnotesize{The Fano plane. The structure constants are determined by the Fano plane, $C_{ijk}=1$ if $ijk$ lies on a line and is ordered according as its orientation. Each oriented line follows the rules of quaternionic multiplication. For example, $e_2e_3=e_5$ and cyclic permutations; odd permutations go against the direction of the arrows on the Fano plane and we pick up a minus sign, e.g. $e_3e_2=-e_5$.}}\label{FANO}
\end{figure}

An element $a\in\Oct$ may be written $a=a^ae_a$, where $a=0,\ldots,7$,  $a^a\in \R$ and $\{e_a\}$ is a basis with one real $e_0$ and  seven $e_i, i=1,\ldots, 7,$ imaginary elements. The octonionic multiplication rule is,
\be
e_ae_b=\left(\delta_{a0}\delta_{bc}+\delta_{0b}\delta_{ac}-\delta_{ab}\delta_{0c}+C_{abc}\right)e_c,
\ee
where $C_{abc}$ is totally antisymmetric and $C_{0bc}=0$.
The non-zero $C_{ijk}$  are given by the Fano plane. See \autoref{FANO}.

The norm preserving algebra is defined as,
\be
\mathfrak{so}(\alg):=\{A\in\Hom_{\R}(\alg) | \langle A a, b\rangle+\langle a, A b\rangle=0, \;\forall a, b \in \alg\}.
\ee
The \emph{triality} algebra of $\alg$ is defined as,
\be\label{TRIDEF}
\mathfrak{tri}(\alg):=\{(A, B, C) \in  \mathfrak{so}(\alg)\oplus\mathfrak{so}(\alg)\oplus\mathfrak{so}(\alg) | A(ab)=B(a)b+aC(b), \; \forall a,b \in \alg\}.\ee
Explicitly,
\be
\begin{split}
\mf{tri}(\R)&\cong \emptyset, \\
\mf{tri}(\C)&\cong\mf{so}(2)\oplus \mf{so}(2),\\
\mf{tri}(\Q)&\cong \mf{so}(3)\oplus \mf{so}(3)\oplus \mf{so}(3),\\
\mf{tri}(\Oct)&\cong  \mf{so}(8).
\end{split}
\ee  

For an element $A\in \mathfrak{so}(\alg)$ define $\tilde{A}\in \mathfrak{so}(\alg)$ by
\be
\tilde{A}(a):=\overline{A(\overline{a})}, \qquad a\in \alg.
\ee
 We can then define an order three Lie algebra automorphism
\be
\theta: \tri(\alg) \rightarrow \tri(\alg) :  (A, B, C)\mapsto (\tilde{B}, C, \tilde{A}),
\ee
which for $\alg=\Oct$ interchanges the three inequivalent 8-dimensional representations of $\mathfrak{so}(\Oct)$.

Given two normed division algebras $\alg_L$ and $\alg_R$ we can define on 
\be\label{eq:triZ2}
\mf{L}_3(\alg_L, \alg_R):=[\mf{tri}(\alg_L)\oplus \mf{tri}(\alg_R)]_{00}+(\alg_L\otimes\alg_R)_{01}+(\alg_L\otimes\alg_R)_{10}+(\alg_L\otimes\alg_R)_{11}
\ee
a $\Z_2\times\Z_2$ graded Lie algebra structure following, with a slight modification to accommodate the required real forms,  Barton and Sudbery \cite{Barton:2003}. First, $\mf{tri}(\alg_L)$ and $\mf{tri}(\alg_R)$ are Lie subalgebras. For 
 elements $T_L=(A_L, \tilde{B}_L, \tilde{C}_L)$
in $\mathfrak{tri}(\alg_L)$
 and 
$(a\otimes b, 0, 0)$, $(0, a\otimes b, 0)$,
and $(0, 0, a\otimes b)$
in $3(\alg_L\otimes\alg_R)$, the commutators are given by the natural action of $\mf{tri}(\alg_L)$,
\be
\begin{split}
[T_L, (a\otimes b, 0, 0)]&=(A_L(a) \otimes b, 0, 0),\\
[T_L, (0, a\otimes b, 0)]&=(0, B_L(a)\otimes b, 0),\\
[T_L, (0, 0, a\otimes b)]&=(0, 0, C_L(a)\otimes b).\\
\end{split}
\ee
Similarly for $T_R=(A_R, \tilde{B}_R, \tilde{C}_R)$
in $\mathfrak{tri}(\alg_R)$,
\be
\begin{split}
[T_R, (a\otimes b, 0, 0)]&=(a \otimes A_R(b), 0, 0),\\
[T_R, (0, a\otimes b, 0)]&=(0, a\otimes B_R(b), 0),\\
[T_R, (0, 0, a\otimes b)]&=(0, 0, a\otimes C_R(b)).\\
\end{split}
\ee
For  two elements belonging to the same summand  $(\alg_L\otimes\alg_R)_{ij}$ in \eqref{eq:triZ2} the commutator is defined using the natural map 
\be
\wedge^2(\alg_L\otimes\alg_R)_i\rightarrow \wedge^2 \alg_L\oplus \wedge^2\alg_R\rightarrow \mf{tri}(\alg_L)\oplus \mf{tri}(\alg_R),
\ee
where the first arrow uses the norm on $\alg_L$ and $\alg_R$. Explicitly, 
\be
\begin{split}
[(a\otimes b, 0, 0), (a'\otimes b', 0, 0)]&=\phantom{-}\blf{a}{a'} \phantom{\theta^2}T^R_{b,b'}+\blf{b}{b'} T^{L}_{a,a'},\\
[(0,a\otimes b, 0), (0, a'\otimes b', 0)]&=-\blf{a}{a'}\phantom{^2}\theta T^R_{b,b'}-\blf{b}{b'}\theta T^{L}_{a,a'},\\
[(0, 0, a\otimes b), (0, 0, a'\otimes b')]&=-\blf{a}{a'}\theta^2 T^R_{b,b'}-\blf{b}{b'}\theta^2 T^{L}_{a,a'}.\\
\end{split}
\ee
Here $T:\wedge^2\alg\rightarrow \tri(\alg):(a,a')\mapsto T_{a, a'}$ is defined by
\be
T_{a,a'}:=(S_{a,a'}, R_{a'}R_{\overline{a}}-R_{a}R_{\overline{a}'},  L_{a'}L_{\overline{a}}-L_{a}L_{\overline{a}'}),
\ee
where
\be
S_{a,a'}(b)=\blf{a}{b}a'-\blf{a'}{b}a,\qquad
L_{a}(b)=ab,\qquad
R_{a}(b)=ba.
\ee
Finally, we have 
\be
\begin{split}
[(a\otimes b, 0, 0), (0, a'\otimes b', 0)]&=\phantom{-}(0, 0, \overline{aa'}\otimes \overline{bb'}),\\
[(0,0,a\otimes b), ( a'\otimes b', 0,0)]&=\phantom{-}(0, \overline{aa'}\otimes \overline{bb'},0),\\
[(0,  a\otimes b, 0), (0, 0, a'\otimes b')]&=-(\overline{aa'}\otimes \overline{bb'},0,0).\\
\end{split}
\ee

With these commutators the magic square formula \eqref{eq:triZ2} describes the Lie algebras of the groups presented in \autoref{tab:ms1}. The simplest way to check that these definitions yield the correct non-compact real forms is by comparison to the conventions of Barton and Sudbery \cite{Barton:2003}, which are known to give the magic square of  compact real forms. Recall, a non-compact real form $\mathfrak{g}_{nc}$ of a complex semi-simple Lie algebra $\mathfrak{g}_{\C}$ admits a symmetric decomposition $\mathfrak{g}_{nc}=\mathfrak{h} + \mathfrak{p}$,
\be
[\mathfrak{h}, \mathfrak{h}]\subseteq \mathfrak{h}, \qquad [\mathfrak{h}, \mathfrak{p}]\subseteq \mathfrak{p},\qquad[\mathfrak{p}, \mathfrak{p}]\subseteq \mathfrak{h},
\ee
where $\mathfrak{h}$ is the maximal compact subalgebra. If a compact real  form $\mathfrak{g}_{c}$ shares with some non-compact  real form $\mathfrak{g}_{nc}$  a common subalgebra,  $\mathfrak{g}_{nc}=\mathfrak{h}+\mathfrak{p}$ and $\mathfrak{g}_{c}=\mathfrak{h}+\mathfrak{p}'$, and the brackets in $[\mathfrak{h}, \mathfrak{p}]$ are the  same  as those in $[\mathfrak{h}, \mathfrak{p}']$, but  equivalent brackets in $[\mathfrak{p}, \mathfrak{p}]$   and $[\mathfrak{p}', \mathfrak{p}']$ differ by a sign, then $\mathfrak{h}$ is the maximal compact subalgebra of $\mathfrak{g}_{nc}$. This observation is sufficient to confirm that our construction yields the real forms in \autoref{tab:ms1} and we can identify 
\be\label{eq:tri2}
\mf{L}_1(\alg_L, \alg_R):=[\mf{tri}(\alg_L)\oplus \mf{tri}(\alg_R)]_{00}+(\alg_L\otimes\alg_R)_{01}
\ee
as the maximal compact subalgebra. The corresponding compact subgroups are presented in \autoref{tab:ms3}.

  \begin{table}[ht]
 \begin{center}
\begin{tabular}{c|ccccccc}
\hline
\hline
 $\alg_L\backslash\alg_R$ && $\R$ & $\C$  & $\Q$  & $\Oct$ & \\
 \hline
 \\
 
   $\R$ && $\SO(2)$ & $\SO(3)\times \SO(2)$   & $\SO(5)\times \SO(3)$   & $\SO(9)$   \\
  $\C$ && $\SO(3)\times \SO(2)$ & $[\SO(3)\times \SO(2)]^2$   & $\SO(6)\times \SO(3)\times \SO(2)$   & $\SO(10)\times \SO(2)$   \\
  $\Q$ && $\SO(5)\times \SO(3)$ & $\SO(6)\times \SO(3)\times \SO(2)$   &$\SO(8)\times \SO(4)$  & $\SO(12)\times \SO(3)$   \\
   $\Oct$ && $\SO(9)$ & $\SO(10)\times \SO(2)$   & $\SO(12)\times \SO(3)$   & $\SO(16)$   \\
   \\
   \hline
   \hline
\end{tabular}
\caption[Magic square of required real forms.]{Magic square of maximal compact subgroups.  \label{tab:ms3}}
 \end{center}
\end{table}

\section{The magic pyramid}\label{sec:mp}

\subsection{Super Yang-Mills over $\RCHO$}

The key to constructing the magic pyramid is to write the appropriate Yang-Mills theories over $\RCHO$ and then consider the symmetries of the ``squared'' theories. The Lagrangian for $(n+2)$-dimensional $\mathcal{N}=1$ SYM with gauge group $G$ over the division algebra $\alg_n$ is given \cite{Baez:2009xt, Anastasiou:2013cya} by
\begin{equation}\label{10-d action}
\mathcal{L}(\Al_n)=-\frac{1}{4}F_{\mu\nu}^AF^{A\mu\nu}-\text{Re}(i\lambda^{\dagger A}\sigmabar^\mu D_\mu\lambda^A),~~~~~\lambda\in\alg_n^2,
\end{equation}
where the covariant derivative and field strength are given by the usual expressions
\be
\begin{split}
D_\mu\lambda^A&=\partial_\mu\lambda^A+gf_{BC}{}^A A_\mu^B\lambda^C,\\
F^A_{\mu\nu}&=\partial_\mu A^A_\nu-\partial_\nu A_\mu^A+gf_{BC}{}^AA^B_\mu A^C_\nu,
\end{split}
\ee
with $A=0,\dots,\dim[G]$. The $\{\sigma^\mu\}$ are a basis for $\Al_n$-valued Hermitian matrices - the straightforward generalisation of the usual complex Pauli matrices \cite{Schray:1994ur, Evans:1994cn, Anastasiou:2013cya} to all four normed division algebras, satisfying the usual Clifford algebra relations. We can use these to write the supersymmetry transformations:
\be\label{SUSYTRANS}
\delta A_\mu^A=\text{Re}(i\lambda^{\dagger A}\sigmabar_\mu \epsilon),\hspace{0.5cm}
\delta\lambda^A=\frac{1}{4}F_{\mu\nu}^A\sigma^\mu(\sigmabar^\nu\epsilon).
\ee
Note, since the components $\lambda^{Aa}$ are anti-commuting we are dealing with the algebra of octonions defined over the Grassmanns.

By dimensionally reducing these theories using the Dixon-halving techniques of \cite{Anastasiou:2013cya}, we arrive at a master Lagrangian for SYM in $D=n+2$ with $\susy$ supersymmetries written over the division algebra $\al_{n\susy}$. The division algebra associated with spacetime $\Al_n$ is viewed as a  subalgebra of $\al_{n\susy}$. The resulting Lagrangian is:
\begin{eqnarray}
\mathcal{L}\left(\Al_n,\Al_{n\susy}\right)=-\frac{1}{4}F^A_{\mu\nu}F^{A\mu\nu}-\frac{1}{2}D_\mu\phi^{A*}D^\mu\phi^A-\text{Re}(i\lambda^{\dagger A}\sigmabar^\mu D_\mu\lambda^A)  ~~~~~~~~~~~~~~~~~~~~~~~~~~~~~~~~~~~~~~~~\hspace{0.07cm}\label{MASTER}\\ 
 \hspace{0.1cm}-gf_{BC}{}^A\text{Re}\left(i{\lambda}^{\dagger A}\varepsilon\phi^B\lambda^C\right)-\frac{1}{16}g^2f_{BC}{}^Af_{DE}{}^A(\phi^{B*}\phi^D+\phi^{D*}\phi^B)(\phi^{C*}\phi^E+\phi^{E*}\phi^C) ,\nonumber
\end{eqnarray}
where $\lambda\in\Al_{n\susy}^2$ (so we have $\susy$ spacetime spinors, each valued in $\Al_{n}^2$) and $\phi$ is a scalar field taking values in $\phi\in\Al_n^\complement$,  the subspace of $\Al_{n\susy}$ orthogonal to the $\al_n$ subalgebra. The $\{\sigmabar^\mu\}$ are still a basis for $\Al_n$-valued Hermitian matrices, again, with $\Al_n$ viewed as a division subalgebra of $\Al_{n\susy}$.  

The supersymmetry transformations in this language are
\be
\begin{split}
\delta A_\mu^A&=\text{Re}(i\lambda^{\dagger A}\sigmabar_\mu \epsilon),\\
\delta\phi^A&=- \frac{i}{2}\tr\left(\varepsilon(\lambda^A\epsilon^\dagger-\epsilon\lambda^{\dagger A})_{\Al_n^\complement}\right) ,\\
\delta\lambda^A&=\frac{1}{4}F_{\mu\nu}^A\sigma^\mu(\sigmabar^\nu\epsilon)+\frac{1}{2}\sigma^\mu\varepsilon(D_\mu\phi^A\epsilon)+\frac{1}{4}f_{BC}{}^A\phi^C(\phi^B\epsilon),
\end{split}
\ee
where the subscript $\Al_n^\complement$ refers to the projection onto this subspace and 
\be
\varepsilon:=\begin{pmatrix}
0 ~ -1 \\
1 ~~~~ 0 \end{pmatrix}.
\ee
The important result is that we can write $\susy=1$ SYM in $D=n+2$ over the division algebra $\Al_n$, and that if we wish to double the amount of supersymmetry we Dickson-double the division algebra. For example, if we start with $\susy=1$ SYM in $D=4$ over $\C$, then $\susy=2$ will be written over $\Q$ and $\susy=4$ will be written over $\Oct$. 

It is also useful to consider the little group representations of the fields in the division algebraic Yang-Mills theories. The little group in $D=n+2$ dimensions is $\SO(n)$. When $\N=1$, SYM contains a single vector and its fermionic partner, which may each be represented as an element of $\Al_n$, with the little group transforming them via the natural action of $\SO(\Al_n)$. For example, in $D=10$ we have an octonion representing the vector, transforming as the $\bold{8}_v$ of SO(8), while the spinor is represented by another octonion, transforming as the $\bold{8}_s$. As noted in \cite{Anastasiou:2013cya}, the overall (little group plus internal) symmetry of the $\N=1$ theory in $D=n+2$ dimensions is $\mf{tri}(\Al_n)$. If we dimensionally reduce these theories we obtain SYM with $\N$ supersymmetries whose overall symmetries are given by
\be
\mf{sym}(\Al_n,\Al_{n\N}):=\big\{(A,B,C)\in \mathfrak{tri}(\Al_{n\N})|~[A, \mf{so}(\Al_n)_{ST}]=0\quad \forall A \notin \mf{so}(\Al_n)_{ST} \big\}
\ee
where  $\mf{so}(\Al_n)_{ST}$ is the subalgbra of $\mathfrak{so}(\Al_{n\N})$ that acts as orthogonal transformations on $\Al_n\subset\Al_{n\N}$. The division algebras used in each dimension and the corresponding $\mf{sym}$ algebras are summarised in Table \ref{YANG}. The on-shell content of each SYM theory can then be summarised as:
\begin{itemize}
\item an $\Al_{n\N}$ element of bosons: a vector in $\Al_n$ and scalars in $\Al_n^\complement$, giving $\Al_n\oplus\Al_n^\complement=\Al_{n\N}$
\item an $\Al_{n\N}$ element of fermions: $\N$ spacetime fermions each valued in $\Al_n$.
\end{itemize}

  \begin{table}[ht]
 \begin{center}
\begin{tabular}{c|ccccccc}
\hline
\hline
 $\Al_n\backslash\Al_{n\susy}$ & $\hspace{1.4cm}\Oct\hspace{1.4cm}$ & $\hspace{1.4cm}\Q\hspace{1.4cm}$  & $\hspace{1.4cm}\C\hspace{1.4cm}$  & $\hspace{1.2cm}\R\hspace{1.0cm}$ & \\
 \hline
 \\
 
   $\Oct$ & $\mf{so}(8)_{ST}$ \\ \\
$\Q$ & $\mf{so}(4)_{ST}\oplus\mf{sp}(1)\oplus\mf{sp}(1)$ & $\mf{so}(4)_{ST}\oplus\mf{sp}(1)$
\\ \\
$\C$ & $\mf{so}(2)_{ST}\oplus\mf{su}(4)$ & $\mf{so}(2)_{ST}\oplus\mf{sp}(1)\oplus\mf{so}(2)$
& $\mf{so}(2)_{ST}\oplus\mf{so}(2)$ \\ \\
$\R$ & $\mf{so}(8)$ & $\mf{so}(4)\oplus\mf{sp}(1)$ & $\mf{so}(2)\oplus\mf{so}(2)$ & $\emptyset$
\\ \\
   \hline
   \hline
\end{tabular}
\caption{\footnotesize{A table of algebras: $\mf{sym}(\Al_n,\Al_{n\N})$. This lets us read off the spacetime and internal symmetries in each Yang-Mills theory. For example, one can see the familiar R-symmetries in $D=4$: U(1), U(2) and SU(4) for $\N=1,2,4$, respectively. Note that the symmetries in $D=3$ are entirely internal and that they include the R-symmetry as a subgroup (these are actually the symmetries of the theories after dualising the vector to a scalar, to be discussed in the following section)}.\label{YANG}}
 \end{center}
\end{table}

\subsection{The $D=3$ magic square of supergravities}

In $D=3$ the spacetime algebra is $\al_1=\R$, so the SYM master Lagrangian (\ref{MASTER}) describes $\N$ real two-component Majorana spinors $\lambda^{\alpha}_{a}$, $\alpha=1, 2$,  written as a single $\al_{\N}$-valued object  $\lambda^{\alpha}$, as well as an $\R$-valued vector and an $\text{Im}\Al_\N$-valued scalar field. Tensoring the multiplets of left ($L$) and right ($R$) Yang-Mills theories, as in \cite{Borsten:2013bp}, we obtain the off-shell field content:
\be\label{eq:sugrafields}
g_{\mu\nu}  \in \R, \quad
\Psi_{\mu}^{\alpha} \in  \begin{pmatrix} \alg_{\N_L}\\ \alg_{\N_R}\end{pmatrix}, \quad
\varphi\in  \begin{pmatrix} \alg_{\N_L}\otimes \alg_{\N_R}\\ \alg_{\N_L}\otimes \alg_{\N_R}\end{pmatrix},\quad \chi^\alpha  \in  \begin{pmatrix} \alg_{\N_L}\otimes \alg_{\N_R}\\ \alg_{\N_L}\otimes \alg_{\N_R}\end{pmatrix}.
\ee
In three dimensions the $\R$-valued graviton and $\alg_{\N_L}\oplus \alg_{\N_R}$-valued gravitino carry no degrees of freedom, but indicate that we have found the field content of a supergravity theory with $\N=\N_L+\N_R$ supersymmetries. 
On-shell the $(\alg_{\N_L}\otimes\alg_{\N_R})^2$-valued scalar and Majorana spinor each have $2({\N_L}\times{\N_R})$ degrees of freedom.

\begin{table}[h]
 \begin{center}
\scriptsize
 \begin{tabular}{c|llllllllllll}
 \hline
 \hline
 \\
$\alg_{\N_L}\backslash\alg_{\N_R}$&&$\R$&$\C$&$\Q$&$\Oct$\\
\\
\hline
\\
&&$\mathcal{N}=2, f=4$&$\mathcal{N}=3, f=8$&$\mathcal{N}=5, f=16$&$\mathcal{N}=9, f=32$\\
$\R$&&$G=\SL(2,\R)$&$G=\SU(2,1)$&$G=\USp(4,2)$&$G=F_{4(-20)}$\\
&&$H=\SO(2)$&$H=\SO(3) \times \SO(2)$&$H=\SO(5)\times \SO(3)$&$H=\SO(9)$\\

&&&&\\

&&$\mathcal{N}=3, f=8$&$\mathcal{N}=4, f=16$&$\mathcal{N}=6, f=32$&$\mathcal{N}=10, f=64$\\
$\C$&&$G=\SU(2,1)$&$G=\SU(2,1)^2$&$G=\SU(4,2)$&$G=E_{6(-14)}$\\
&&$H=\SO(3) \times \SO(2)$&$H=\SO(3)^2 \times \SO(2)^2$&$H=\SO(6) \times \SO(3)\times \SO(2)$&$H=\SO(10) \times \SO(2)$\\

&&&\\

&&$\mathcal{N}=5, f=16$&$\mathcal{N}=6, f=32$&$\mathcal{N}=8, f=64$&$\mathcal{N}=12, f=128$\\
$\Q$&&$G=\USp(4,2)$&$G=\SU(4,2)$&$G=\SO(8,4)$&$G=E_{7(-5)}$\\
&&$H=\SO(5)\times \SO(3)$&$H=\SO(6)\times \SO(3)\times \SO(2)$&$H=\SO(8)\times \SO(3)\times \SO(3)$&$H=\SO(12)\times \SO(3)$\\

&&&&\\

&&$\mathcal{N}=9, f=32$&$\mathcal{N}=10, f=64$&$\mathcal{N}=12, f=128$&$\mathcal{N}=16, f=256$\\
$\Oct$&&$G=F_{4(-20)}$&$G=E_{6(-14)}$&$G=E_{7(-5)}$&$G=E_{8(8)}$\\
&&$H=\SO(9)$&$H=\SO(10) \times \SO(2)$&$H=\SO(12)\times \SO(3)$&$H=\SO(16)$\\
\\
\hline
\hline
\end{tabular}
\caption{\footnotesize{Pyramid base   ($D=3$ supergravity).  The first row of each entry indicates the amount of supersymmetry $\mathcal{N}$ and the total number of degrees of freedom $f$. The second (third) row indicates the U-duality group $G$ (the maximal compact subgroup $H\subset G$) and its dimension. The scalar fields in each case parametrise the coset $G/H$, where $\dim_\R(G/H)=f/2$. The } }.
\label{tab:3Dsugra}
  \end{center}
\end{table}

The maximal compact subgroups $H$ of the magic square of U-dualities $G$ are those given in the reduced magic square presented in \autoref{tab:ms3}. These are the largest linearly-realised global symmetries under which   $\Psi_\mu, \varphi$ and $\chi$ transform as a vector, spinor and conjugate spinor;  $\alg_{\mathcal{N}_L}\oplus \alg_{\mathcal{N}_R}$ and $(\alg_{\mathcal{N}_L}\otimes\alg_{\mathcal{N}_R})^2$ are precisely the representation spaces of the vector and (conjugate) spinor. For example, in the maximal case of $\alg_{\mathcal{N}_L}, \alg_{\mathcal{N}_R}=\Oct$, the U-duality is $G=E_{8(8)}$ and we have the $\mathbf{16}, \mathbf{128}$ and $\mathbf{128'}$ of its maximal compact subgroup $H=\SO(16)$.  

We can better understand the origin of the magic $D=3$ U-dualities if we consider the symmetries of their parent Yang-Mills theories. If we set the coupling constant $g$ in (\ref{MASTER}) to zero then we may dualise the vector to a scalar and write the Lagrangian as:
\be\label{eq:dualLD3}
\mathcal{L}(\al_{\susy})=-\frac{1}{2}\partial_\mu\phi^{A*}\partial^\mu\phi^A-i\lambda^{\dagger A}\sigmabar^\mu \partial_\mu\lambda^A,
\ee
where $\phi$ and $\lambda^{\alpha}$ each take values in $\Al_\N$ (note that in $D=3$ we do not need to take the real part of the fermion kinetic term since the sigma matrices are real and symmetric). The supersymmetry transformations become:
\be\label{DUALISEDSUSY}
\delta\phi^A=-i\epsilon^\dagger\varepsilon\lambda^A,~~~~~~\delta\lambda^A=\frac{1}{2}\sigma^\mu\varepsilon\partial_\mu\phi^A\epsilon.
\ee
Equation \eqref{eq:dualLD3} and \eqref{DUALISEDSUSY}   enjoy a symmetry whose Lie algebra is  $\mf{sym}(\R,\Al_{\susy})\cong\mathfrak{tri}(\Al_\N)$. Looking at the theory it is clear why this is the case, since transformations preserving  (\ref{DUALISEDSUSY})  coincide with the definition of $\mathfrak{tri}(\Al_\N)$; that is to say, we might initially try to transform the three $\Al_\N$-valued objects $\phi$, $\lambda$ and $\epsilon$ by three independent $\SO(\Al_\N)$ rotations, but the supersymmetry transformations constrain these three rotations to satisfy the definition (\ref{TRIDEF}).

Tensoring the on-shell field content, $\phi_{L/R}, \lambda_{L/R}$, of the dualised Yang-Mills theories we obtain an $(\alg_{\N_L}\otimes\alg_{\N_R})^2$-valued scalar and  spinor,
\be\label{eq:sugrafieldsonshell}
\varphi\in  \begin{pmatrix} \phi_L\otimes \phi_R\\ \lambda_L\otimes \lambda_R \end{pmatrix},\quad \chi  \in  \begin{pmatrix} \phi_L\otimes \lambda_R\\\lambda_L\otimes \phi_R\end{pmatrix}.
\ee
The $\mathfrak{tri}(\Al_{\N_L})$ and $\mathfrak{tri}(\Al_{\N_R})$ symmetries of the two Yang-Mills theories act on these doublets as the (operator valued) $2\times 2$ matrices:
\be
\begin{pmatrix}\mathfrak{tri}(\Al_{\N_L})\otimes\id & 0\\ 0& \mathfrak{tri}(\Al_{\N_L})\otimes\id \end{pmatrix},~~~\begin{pmatrix} \id\otimes~\mathfrak{tri}(\Al_{\N_R})& 0\\ 0& \id\otimes~\mathfrak{tri}(\Al_{\N_R})\end{pmatrix}.
\ee
Since we have arranged the fields of the squared theory into doublets we might also consider the off-diagonal rotations
\be\label{eq:offdiagH}
\begin{pmatrix} 0& e^*_a\ox e^*_{a'}\\ -e_a\ox e_{a'}& 0\end{pmatrix}.
\ee
The total algebra of linear transformations is then given by
\be
\mf{L}_1(\alg_L, \alg_R):=\mf{tri}(\Al_{\N_L})\oplus \mf{tri}(\Al_{\N_R})+\Al_{\N_L}\otimes\Al_{\N_R},
\ee
which is precisely that of the reduced magic square. It is interesting to note that the off-diagonal transformations of \eqref{eq:offdiagH} take Yang-Mills fermions into Yang-Mills bosons, and vice versa, but are bosonic generators in the supergravity theory. It is tempting to identify these generators with $Q_L\otimes Q_R$, where $Q_{L/R}$ are the supercharges of the left and right Yang-Mills theories, as this correctly reproduces basic structure of transformations \eqref{eq:offdiagH}. However, the derivative in the supersymmetry transformations appears to spoil this correspondence.

The full non-linear U-duality groups $G$ are fixed by the field content and $H$ symmetries, as described in \cite{deWit:1992up}. The groups we find are, of course, those of the magic square in \autoref{tab:ms1}. We summarise the theories of the $D=3$ magic square in \autoref{tab:3Dsugra}.

\subsection{Generalisation to $D=4,6,10$}

To generalise the results above to $D=4,6,10$ we first consider the simplest example in $D=4$: tensoring two $\N=1$ SYM multiplets $(A_\mu,\lambda)$ to obtain $\N=2$ supergravity. Counting the degrees of freedom ($4\times 4=16$) tells us that we must have a gravity multiplet coupled to one hypermultiplet, so the field content we expect is: $(g_{\mu\nu},2\Psi_\mu,A_\mu,2\chi,4\phi)$. We will square on-shell, so the Yang-Mills fields are represented by the complex numbers (helicity states):
\be
A_{L/R},~\lambda_{L/R}\in\C_{L/R}.
\ee
Squaring and arranging into doublets of bosons and fermions then gives us the $(\C\ox\C)^2$ valued objects:
\be
B=\begin{pmatrix} A_L\ox A_R \\ \lambda_L\ox\lambda_R\end{pmatrix}~~\text{and}~~~
F=\begin{pmatrix} A_L\ox\lambda_R \\ \lambda_L\ox A_R\end{pmatrix}.
\ee
Consider acting on these with $\mf{L}_1(\C_L, \C_R):=\mf{tri}(\C_L)\oplus \mf{tri}(\C_R)+\C_L\otimes\C_R$. A basis for $\tri(\C_L)\cong \mathfrak{so}(2)\oplus\mathfrak{so}(2)$ is given by the matrices
\be
\begin{pmatrix} i\ox 1 & 0 \\ 0 & i\ox 1\end{pmatrix}, ~~~\begin{pmatrix} i\ox 1 & 0 \\ 0 &- i\ox 1\end{pmatrix},
\ee
while $\tri(\C_R)$ has
\be
\begin{pmatrix} 1\ox i & 0 \\ 0 & 1\ox i\end{pmatrix}, ~~~\begin{pmatrix} 1\ox i & 0 \\ 0 &- 1\ox i\end{pmatrix},
\ee
and the $\C_L\ox\C_R$ part consists of the four anti-Hermitian matrices
\be
\begin{pmatrix} 0 & 1\ox 1\\ -1\ox 1 & 0\end{pmatrix}, 
~~~\begin{pmatrix} 0 & -i\ox 1\\ -i\ox 1 & 0\end{pmatrix},
~~~\begin{pmatrix} 0 & -1\ox i\\ -1\ox i & 0\end{pmatrix},
~~~\begin{pmatrix} 0 & i\ox i\\ -i\ox i & 0\end{pmatrix}.
\ee
When working with $\C\ox\C$ it is convenient to define the quantities
\be
\begin{split}
1_{\pm}:=\frac{1}{2}(1\ox1\mp i\ox i),~~~~i_{\pm}:=\frac{1}{2}(i\ox 1\pm 1\ox i),
\end{split}
\ee
which each seperately behave like the basis of $\C$:
\be
\begin{split}
i_{\pm}^2=-1_{\pm},~~~~1_{\pm}i_{\pm}=i_{\pm},~~~~1_{\pm}^2=1_{\pm},
\end{split}
\ee
but annihilate one another\footnote{The objects $1_{\pm}$ act as projection operators dividing $\C\ox\C$ into two 2-dimensional subspaces, on which $i_{\pm}$ act as complex structures, so that $\C\ox\C\cong\C\oplus\C$.}:
\be
1_{\pm}1_{\mp}=0,~~~~1_{\pm}i_{\mp}=0,~~~~i_{\pm}i_{\mp}=0.
\ee
Rewriting the 8 matrices above in terms of this basis we obtain the following set:
\be
\begin{split}
&i_+\id, \; i_+\sigma^1,\;1_+\eps,\; i_+\sigma^3,\\
&i_-\id,\;i_-\sigma^1,\;1_-\eps,\;i_-\sigma^3,
\end{split}
\ee
where the sigmas and epsilon refer to the usual Pauli matrices (as opposed to the generalised Pauli matrices defined previously): 
\be
\sigma^1=\begin{pmatrix}0&1\\1&0\end{pmatrix},~~~~~\eps=\begin{pmatrix}0&-1\\1&0\end{pmatrix},
~~~~~\sigma^3=\begin{pmatrix}1&0\\0&-1\end{pmatrix}.
\ee
This set of matrices evidently generates $\SU(2)\times\SU(2)\times\Un(1)\times\Un(1)\cong\SO(4)\times\SO(2)\times\SO(2)$, as stated in the magic square in \autoref{tab:ms3}.

However, the original Yang-Mills theories have a symmetry under the transformations
\be
\delta A_{L/R}=i\theta_{ST} A_{L/R},~~~~~\delta\psi_{L/R}=\frac{1}{2}i\theta_{ST}\psi_{L/R} +i\theta^{L/R}_{I}\psi_{L/R},
\ee
where $\theta_{ST}$ is the spacetime $\Un(1)_{ST}$ little group parameter and $\theta^{L/R}_I$ are those of the Yang-Mills R-symmetries. Note that these correspond to $\tri(\C_{L/R})\cong \mathfrak{so}(2)\oplus\mathfrak{so}(2)$. The variation of the fermionic doublet under $\Un(1)_{ST}$ and the R-symmetries is thus
\be
\delta F=\left[\theta_{ST}\begin{pmatrix} i\ox 1~~~~0~~ \\~~~ 0~~~\frac{1}{2}i\ox 1\end{pmatrix}\hspace{-0.05cm}+\hspace{-0.05cm}
\theta_{ST}\begin{pmatrix}\frac{1}{2}1\ox i ~~~0~~~ \\~~~ 0~~~1\ox i \end{pmatrix}
\hspace{-0.05cm}+\hspace{-0.05cm}
\theta^{L}_{I}\begin{pmatrix}0 ~~~~~0~~ \\ 0~~ i\ox 1\end{pmatrix}\hspace{-0.05cm}+\hspace{-0.05cm}
\theta^{R}_{I}\begin{pmatrix}1\ox i &0 \\ 0&0\end{pmatrix}\right]
F.
\ee
Written in terms of the $i_{\pm}$ basis  we find
\be
\delta F=\left[\theta_{ST}\left(\frac{3}{2}i_+\id+\frac{1}{2}i_-\sigma^3\right)+\frac{1}{2}\theta_{I}^L(\id-\sigma^3)(i_++i_-)+\frac{1}{2}\theta_{I}^R(\id+\sigma^3)(i_+-i_-) \right]F.
\ee
What has emerged are the $\Un(1)_{ST}$ transformations for the gravitinos and spin-$\frac{1}{2}$ fermions and two internal U(1) pieces. Since $\C\ox\C$ is not a division algebra, it contains zero divisors; it is interesting to note their role here in ensuring that the spin-$\frac{3}{2}$ and spin-$\frac{1}{2}$ fields each receive their appropriate little group transformations. Now the largest symmetry that acts on this doublet must be the subalgebra of $\mf{L}_1(\C_L, \C_R)$ that commutes with these spacetime generators. All the matrices commute with $i_+\id$, but  $i_-\sigma^1$ and $1_-\eps$ do not commute with $i_-\sigma^3$, so we are forced to discard these generators. The remaining matrices generate:
\be
\Un(1)_{ST}\times\left(\Un(1)\times\Un(1)\times\SU(2)\right)_U,
\ee 
where the subscript $U$ denotes the maximal compact subgroup of the U-duality. This is the entry found in the pyramid. Again, note how the gravitino transforms as a doublet under the $\SU(2)$ but, because of the $i_+$, the spin-$\frac{1}{2}$ fields are singlets, as required in the supergravity theory. The Yang-Mills R-symmetries have been absorbed into the U-duality group.  A similar analysis for the bosonic fields in the theory shows that we do indeed obtain a graviton, a vector and two scalars, which transform as a singlet, a singlet and doublet under the $\SU(2)$, just as one would hope.

In the general case for $D=n+2$, we begin with a pair of Yang-Mills theories with $\N_L$ and $\N_R$ supersymmetries written over the division algebras $\Al_{n\N_L}$ and $\Al_{n\N_R}$, respectively. Taking the little group fields we may then write all the bosons of the left (right) theory as a single element $B_L\in \Al_{n\N_L}$ ($B_R\in \Al_{n\N_R}$), and similarly for the fermions $F_L\in \Al_{n\N_L}$ ($F_R\in \Al_{n\N_R}$). After tensoring we arrange the resulting supergravity fields into a bosonic doublet and a fermionic doublet, 
\be\label{eq:sugrafieldsonshellD}
B= \begin{pmatrix} B_L\otimes B_R\\ F_L\otimes F_R \end{pmatrix},\quad F  =  \begin{pmatrix} B_L\otimes F_R\\F_L\otimes B_R\end{pmatrix},
\ee
just as we did in $D=3$. The largest algebra that can act on these doublets is $\mf{L}_1(\Al_{n\N_L},\Al_{n\N_R})$, but an $\mf{so}(\al_n)_{ST}$ subalgebra of this corresponds to spacetime transformations, so we must restrict $\mf{L}_1(\Al_{n\N_L},\Al_{n\N_R})$ to the subalgebra that commutes with $\mf{so}(\al_n)_{ST}$.

Similarly, for the full non-compact groups $G$ it is, of course, a necessary condition that the required subalgebra of $\mf{L}_3(\Al_{n\N_L},\Al_{n\N_R})$ commutes with $\mf{so}(\al_n)_{ST}$. Imposing this condition actually turns out to be sufficient to find all the correct U-dualities. The Lie algebra $\mf{u}$ of the U-duality group of a $D=n+2$ supergravity theory obtained by tensoring Yang-Mills theories with $\N_L$ and $\N_R$ supersymmetries is thus given by:
\be\label{eq:pyramid}
\mf{u}\cong\mathfrak{Pyr}(\alg_{n}, \alg_{n\susy_L}, \alg_{n\susy_R}):=\left\{u\in\mf{L}_3(\alg_{n\N_L}, \alg_{n\N_R})-\mf{so}(\alg_n)_{ST}\Big|[u,\mf{so}(\alg_n)_{ST}]=0\right\}.
\ee
To evaluate this formula for different values of $\N_L$, $\N_R$ and $n$, we require only to decompose the adjoint representations of groups given by $\mf{L}_3(\alg_{n\N_L}, \alg_{n\N_R})$ under the $\SO(\alg_n)_{ST}$ subgroup and discard all pieces that transform non-trivially under $\SO(\alg_n)_{ST}$. 

Locating the spacetime subgroup in the supergravity just amounts to tracking it back to the Yang-Mills theories by decomposing $\mf{tri}(\Al_{n\N_{L/R}})$ into $\mf{sym}(\Al_n,\Al_{n\N_{L/R}})$. The systematic process for finding the U-dualities can then be summarised by the following recipe:
\begin{itemize}
\item Decompose $\mf{tri}(\Al_{n\N_{L}})\oplus\mf{tri}(\Al_{n\N_{R}})$ into $\mf{sym}(\Al_n,\Al_{n\N_{L}})\oplus\mf{sym}(\Al_n,\Al_{n\N_{R}})$,
\item Identify $\SO(\Al_n)_{ST}$ as the diagonal subgroup in $\SO(\Al_n)_{ST_L}\times\SO(\Al_n)_{ST_R}$,
\item Discard all generators that transform non-trivially under $\SO(\Al_n)_{ST}$.
\end{itemize}
For the maximal compact subgroups $H$, we just follow the same recipe with $\mf{L}_1(\alg_{n\N_L}, \alg_{n\N_R})$. To extract the spacetime representations contained in the doublets $B,F$ we decompose the spinor and conjugate spinor representations of $\mf{L}_1(\alg_{n\N_L}, \alg_{n\N_R})$ with respect to $H\times\SO(\Al_n)_{ST}$. In the following we summarise each layer, demonstrating the calculation of the U-duality and tabulating the resulting supergravities.

\paragraph{$D=4$ layer:}
As the archetypal example, consider squaring $D=4$, $\N=4$ Yang-Mills to obtain $\N=8$ supergravity, which we know should have $E_{7(7)}$ as its U-duality, with a maximal compact subgroup of SU(8). Following the recipe above we decompose
\be
\begin{split}
E_{8(8)}\supset \SO(8)\times\SO(&8)\supset\SU(4)\times\Un(1)\times\SU(4)\times\Un(1)\\
\bold{248}\rightarrow\,\,\phantom{+}\big[\bold{(15,1)}_{00}&+\bold{(1,1)}_{00}\,\,\,\,+\bold{(1,15)}_{00}+\bold{(1,1)}_{00}\\
+\bold{(6,1)}_{20}&+\bold{(1,6)}_{02}\,\,\,\,+\bold{(1,6)}_{0-2}+\bold{(6,1)}_{-20}\\
+\bold{(4,4)}_{11}&+\bold{(4,\overline{4})}_{1-1}+\bold{(\overline{4}, 4)}_{-11}+\bold{(\overline{4}, \overline{4})}_{-1-1}\big]\\
\big[+\bold{(\overline{4}, \overline{4})}_{11}&+\bold{(\overline{4}, 4)}_{1-1}+\bold{(4,\overline{4})}_{-11}+\bold{(4,4)}_{-1-1}\\
+\bold{(1,1)}_{22}&+\bold{(1,1)}_{2-2}+\bold{(1,1)}_{-22}+\bold{(1,1)}_{-2-2}\\
+\bold{(1,6)}_{20}&+\bold{(6,1)}_{02}\,\,\,\,+\bold{(6,1)}_{0-2}+\bold{(1,6)}_{-20}+\bold{(6,6)}_{00}\big],
\end{split}
\ee
where we have split the compact and non-compact generators with the square brackets. Adding and subtracting the U(1) charges, the first is the charge under $\Un(1)_{ST}$, while the second is an internal charge:
\be
\begin{split}
E_{8(8)}\supset \SO(8)\times\SO(&8)\supset\SU(4)\times\Un(1)\times\SU(4)\times\Un(1)\\
\bold{248}\rightarrow\,\,\phantom{+}\big[\bold{(15,1)}_{00}&+\bold{(1,1)}_{00}\hspace{0.24cm}+\bold{(1,15)}_{00}+\bold{(1,1)}_{00}\\
+\bold{(6,1)}_{22}&+\bold{(1,6)}_{2-2}+\bold{(1,6)}_{-22}+\bold{(6,1)}_{-2-2}\\
+\bold{(4,4)}_{20}&+\bold{(4,\overline{4})}_{02}\hspace{0.24cm}+\bold{(\overline{4}, 4)}_{0-2}+\bold{(\overline{4}, \overline{4})}_{-20}\big]\\
\big[+\bold{(\overline{4}, \overline{4})}_{20}&+\bold{(\overline{4}, 4)}_{02}\hspace{0.24cm}+\bold{(4,\overline{4})}_{0-2}+\bold{(4,4)}_{-20}\\
+\bold{(1,1)}_{40}&+\bold{(1,1)}_{04}\hspace{0.24cm}+\bold{(1,1)}_{0-4}+\bold{(1,1)}_{-40}\\
+\bold{(1,6)}_{22}&+\bold{(6,1)}_{2-2}+\bold{(6,1)}_{-22}+\bold{(1,6)}_{-2-2}+\bold{(6,6)}_{00}\big],
\end{split}
\ee
Discarding those generators carrying non-trivial  $\Un(1)_{ST}$ charge along with  $\Un(1)_{ST}$ itself, we recognise the decomposition of $E_{7(7)}$ into $\SU(4)\times\SU(4)\times\Un(1)$:
\be
\begin{split}
\bold{133}\rightarrow&\,\,\phantom{+}\big[\bold{(1,1)}_{0}+\bold{(15,1)}_{0}+\bold{(1,15)}_{0}+\bold{(4,\overline{4})}_{2}+\bold{(\overline{4}, 4)}_{-2}\big]\\
&+\big[\bold{(4,\overline{4})}_{-2}+\bold{(\overline{4}, 4)}_{2}+\bold{(1,1)}_{4}+\bold{(1,1)}_{-4}+\bold{(6,6)}_{0}\big],
\end{split}
\ee
where we have suppressed the $\Un(1)_{ST}$  spacetime charges, which are all zero. The compact pieces form the maximal compact subgroup $\SU(8)$:
\be
\begin{split}
\SU(8)\supset&\SU(4)\times\SU(4)\times\Un(1)\\
\bold{63}\rightarrow&\,\,\bold{(1,1)}_{0}+\bold{(15,1)}_{0}+\bold{(1,15)}_{0}+\bold{(4,\overline{4})}_{2}+\bold{(\overline{4}, 4)}_{-2}.
\end{split}
\ee
To extract the field content we simply  decompose the $\bold{128}$ ($B$) and $\bold{128}'$ ($F$) of SO(16) with respect to $\SU(8)\times\Un(1)_{ST}$:
\be
\begin{split}
\bold{128}\hspace{0.1cm}&\rightarrow\bold{1}_{4}+\bold{1}_{-4}+\bold{28}_{2}+\overline{\bold{28}}_{-2}+\bold{70}_{0}\\
\bold{128}'&\rightarrow\bold{8}_{3}+\overline{\bold{8}}_{-3}+\bold{56}_{1}+\overline{\bold{56}}_{-1},
\end{split}
\ee
 which yields the expected supermultiplet\footnote{Further branching the SU(8) representations above with respect to $\SU(4)\times\SU(4)$ we can see their Yang-Mills origins more clearly.}: $(g_{\mu\nu},8\Psi_\mu,28A_\mu,56\chi,70\phi)$. Repeating this process for the other theories in $D=4$ gives \autoref{tab:4Dsugra}. These theories were previously obtained in \cite{Carrasco:2012ca} by consistently truncating to the untwisted sector of the low-energy effective field theory  of type II superstrings on factorised orbifolds,  revealing their double-copy structure. The magic $D=4$, $\mathcal{N}=2$ supergravities were also discussed in this context. In particular, the quarternionic theory originates from a non-factorisable $\Z_2$-orbifold compactification \cite{Carrasco:2012ca}. 

\begin{table}[h]
 \begin{center}
\scriptsize
 \begin{tabular}{c|llllllllllll}
 \hline
 \hline
 \\
$\alg_{2\N_L}\backslash \alg_{2\N_R}$&&$\C$&$\Q$&$\Oct$\\
\\
\hline
\\
&&$\mathcal{N}=2, f=16$&$\mathcal{N}=3, f=32$&$\mathcal{N}=5, f=64$\\
$\C$&&$G=\Un(1,2)$&$G=\Un(1,3)$&$G=\SU(1,5)$\\
&&$H=\Un(1)\times \Un(2)$&$H=\Un(1)\times\Un(3) $&$H=\Un(5)$\\

&&&&\\

&&$\mathcal{N}=3, f=32$&$\mathcal{N}=4, f=64$&$\mathcal{N}=6, f=128$\\
$\Q$&&$G=\Un(1,3)$&$G=\SL(2, \R)\times \SO(6,2)$&$G=\SO^\star(12)$\\
&&$H=\Un(1)\times\Un(3)$&$H=\Un(1)\times\Un(4)$&$H=\Un(6)$\\

&&&\\

&&$\mathcal{N}=5, f=64$&$\mathcal{N}=6, f=128$&$\mathcal{N}=8, f=256$\\
$\Oct$&&$G=\SU(1,5)$&$G=\SO^\star(12)$&$G=E_{7(7)}$\\
&&$H=\Un(5)$&$H=\Un(6)$&$H=\SU(8)$\\
\\
\hline
\hline
\end{tabular}
\caption{\footnotesize{First floor  of pyramid ($D=4$ supergravity).  The first row of each entry indicates the amount of supersymmetry $\mathcal{N}$ and the total number of degrees of freedom $f$. The second (third) row indicates the U-duality group $G$ (the maximal compact subgroup $H\subset G$) and its dimension. The scalar fields in each case parametrise the coset $G/H$.}}.
\label{tab:4Dsugra}
  \end{center}
\end{table}

\FloatBarrier

\paragraph{$D=6$ layer:}

In $D=6$ the spacetime little group is $\SO(4)_{ST}\cong\Sp(1)^+_{ST}\times\Sp(1)^-_{ST}$, and the Yang-Mills multiplets available are $(\mathcal{N}_+, \mathcal{N}_-)=(1, 0),(0,1),(1,1)$, written over $\Q,\Q,\Oct$, respectively. Here we summarise the tensoring of these multiplets. For 
\be
[(\mathcal{N}_+, \mathcal{N}_-)_{SYM}^L]\times [({\mathcal{N}}_+, {\mathcal{N}}_-)_{SYM}^R]
\ee
we have:
\be
\begin{split}
 [(1,1)_{SYM}^{L}]\times[(1,1)_{SYM}^{R}]& = [(2,2)_{sugra}]\\
 [(1,1)_{SYM}^{L}]\times[(0,1)_{SYM}^{R}] &= [(1,2)_{sugra}]\\
  [(1,0)_{SYM}^{L}]\times[(1,0)_{SYM}^{R}] &= [(2,0)_{sugra}+(2,0)_{tensor}]
\end{split}
\ee
The details of the above tensorings are given in \autoref{sec:6dtensor}. See also \cite{deWit:2002vz}. The chiral $(1, 2)_{sugra}$ is anomalous and adding the required compensating matter extends the theory to $(2,2)_{sugra}$ \cite{DAuria:1997cz}.  Similarly, the  $[(2,0)_{sugra}+(2,0)_{tensor}]$ theory is anomalous since the unique anomaly free $\mathcal{N}=(2,0)$ theory consists of one $(2,0)_{sugra}$ multiplet coupled to 21 $(2,0)_{tensor}$ multiplets as obtained by compactifying $D=10$ Type IIB supergravity on a $K3$.  The additional 20  $(2,0)_{tensor}$ multiplets required to cancel the anomaly can be included by considering an alternative tensoring with $\mathcal{N}_L=(2,0)$ and $\mathcal{N}_R=(0,0)$,
\be
[(2,0)_{tensor}^{L}]\times[B_{\mu\nu}^{- R}+21\phi^R] = [(2,0)_{sugra}+21 (2,0)_{tensor}],
\ee
where $B_{\mu\nu}^{- R}$ is  anti-selfdual and transforms as a $(\rep{1, 3})$ under the space-time little group.

Note, we could have also chosen the left/right $\mathcal{N}=1$ multiplets to have opposite chiralities yielding,
\be
 [(1,0)_{SYM}^{L}]\times[(0,1)_{SYM}^{R}] = [(1,1)_{sugra}].
\ee
To demonstrate the calculation of the U-duality groups, consider the maximal case, $[(1,1)_{SYM}^{L}]\times[(1,1)_{SYM}^{R}]= [(2,2)_{sugra}]$. Following the recipe gives
\be
\begin{split}
E_{8(8)}\supset \SO(8)\times\SO(8)\supset\Sp(&1)^+_{ST_L}\times\Sp(1)^-_{ST_L} \times\Sp(1)\times\Sp(1)\times\Sp(1)^+_{ST_R}\times\Sp(1)^-_{ST_R} \times\Sp(1)\times\Sp(1)\\
\mathbf{248} \to\phantom{+\!}\big[\mathbf{(3,1,1,1,1,1,1,1)}&\mathbf{+(1,3,1,1,1,1,1,1)+(1,1,3,1,1,1,1,1)+(1,1,1,3,1,1,1,1)} \\
\phantom{\to}+\mathbf{(1,1,1,1,3,1,1,1)}&\mathbf{+(1,1,1,1,1,3,1,1)+(1,1,1,1,1,1,3,1)+(1,1,1,1,1,1,1,3)}\\
\phantom{\to}+\mathbf{(1,2,2,1,1,2,2,1)}&\mathbf{+(1,2,2,1,2,1,1,2)+(2,1,1,2,1,2,2,1)+(2,1,1,2,2,1,1,2)} \\
\phantom{\to}+\mathbf{(2,2,2,2,1,1,1,1)}&\mathbf{+(1,1,1,1,2,2,2,2)}\big]\\
\phantom{\to}+\big[\mathbf{(2,2,1,1,2,2,1,1)}&\mathbf{+(2,2,1,1,1,1,2,2)+(1,1,2,2,2,2,1,1)+(1,1,2,2,1,1,2,2)} \\
\phantom{\to}+\mathbf{(2,1,2,1,2,1,2,1)}&\mathbf{+(2,1,2,1,1,2,1,2)+(1,2,1,2,2,1,2,1)+(1,2,1,2,1,2,1,2)} \big],
\end{split}
\ee
where, as before, the square brackets partition the generators into those that live in the maximal compact subgroup and those that do not. Once again, it appears that we have two copies of the spacetime little group; we must take the diagonal subgroup: $\Sp(1)^+_{ST_L}\times\Sp(1)^-_{ST_L}\times\Sp(1)^+_{ST_R}\times\Sp(1)^-_{ST_R} \rightarrow\Sp(1)^+_{ST}\times\Sp(1)^-_{ST}$. This means we tensor product the representations appearing in the corresponding slots (that is, we identify the first slot with the fifth and identify the second with the sixth), leading to:
\be\begin{split}
E_{8(8)}\supset \Sp(1)^+_{ST}\times\Sp(1)^-_{ST} &\times\Sp(1)\times\Sp(1)\times\Sp(1)\times\Sp(1)\\
\mathbf{248} \to\phantom{+\!}\big[\mathbf{(3,1,1,1,1,1)}&\mathbf{+(3,1,1,1,1,1)+(1,3,1,1,1,1)+(1,3,1,1,1,1)} \\
\phantom{\to}+\mathbf{(1,1,3,1,1,1)}&\mathbf{+(1,1,1,3,1,1)+(1,1,1,1,3,1)+(1,1,1,1,1,3)} \\
\phantom{\to}+\mathbf{(1,3,2,1,2,1)}&\mathbf{+(1,1,2,1,2,1)+(2,2,2,1,1,2)+(2,2,1,2,2,1)} \\
\phantom{\to}+\mathbf{(3,1,1,2,1,2)}&\mathbf{+(1,1,1,2,1,2)+(2,2,2,2,1,1)+(2,2,1,1,2,2)}\big]\\
\phantom{\to}+\big[\mathbf{(3,3,1,1,1,1)}&\mathbf{+(1,3,1,1,1,1)+(3,1,1,1,1,1)+(1,1,1,1,1,1)}\\
\phantom{\to}+\mathbf{(2,2,1,1,2,2)}&\mathbf{+(2,2,2,2,1,1)+(1,1,2,2,2,2)} \\
\phantom{\to}+\mathbf{(3,1,2,1,2,1)}&\mathbf{+(1,1,2,1,2,1)+(2,2,2,1,1,2)}\\
\phantom{\to}+\mathbf{(2,2,1,2,2,1)}&\mathbf{+(1,3,1,2,1,2)+(1,1,1,2,1,2)} \big],
\end{split}
\ee
where the first and second slots label $\Sp(1)^+_{ST}\times\Sp(1)^-_{ST}$. Truncating all pieces that are not spacetime singlets we find the remaining generators are those of the following decomposition:
\be\begin{split}
\SO(5,5)&\supset\SO(5)\times\SO(5)\supset\Sp(1)\times\Sp(1)\times\Sp(1)\times\Sp(1)\\
\mathbf{45} &\to\phantom{+\!}\mathbf{\big[(3,1,1,1)+(1,3,1,1)+(1,1,3,1)+(1,1,1,3)+(2,1,2,1)+(1,2,1,2)\big]} \\
&\phantom{\to}+\mathbf{\big[(2,2,2,2)+(2,1,2,1)+(1,2,1,2)+(1,1,1,1)\big]}.
\end{split}
\ee
Note that the generators in the first pair of square brackets belong to the maximal compact subgroup $\SO(5)\times\SO(5)$, and those in the second pair are all non-compact, so we do indeed find the maximally non-compact real form $\SO(5,5)$, familiar from the dimensional reduction of Type II supergravity on $T^4$.

Applying this procedure to the other two slots in $D=6$ we recover \autoref{tab:6Dsugra}, where we have chosen to tensor SYM mutiplets of opposite chiralities in the $\Q\ox\Q$ case, resulting in pure $\mathcal{N}=(1,1)$ supergravity with $G/H$ given by $\Sp(1)\times \Sp(1)\times \textrm{O}(1,1)/\Sp(1)\times\Sp(1)$. On the other hand, for matching chiralities we obtain   $\mathcal{N}=(2, 0)$ supergravity coupled to a single tensor multiplet with  $G/H$ given by $\Sp(2)\times \textrm{O}(1,1)/\Sp(2)$.

Although we do not consider them directly here, it should be noted that the magic $D=6$, $\mathcal{N}=(1,0)$  supergravities (which come coupled to $2,3,5,9$ tensor multiplets and $2,4,8,16$ vector multiplets, respectively,  as well as hypers) are closely related to the magic square and constitute the parent theories of the magic $D=5,4,3$ supergavities. See   \cite{Gunaydin:2010fi} and the references therein.

\begin{table}
 \begin{center}
\scriptsize
 \begin{tabular}{c|llllllllllll}
 \hline
 \hline
 \\
$\alg_{4\N_L}\backslash\alg_{4\N_R}$&&$\Q$&$\Oct$\\
\\
\hline
\\
&&$\mathcal{N}=(1,1), f=64$&$\mathcal{N}=(1, 2), f=128$\\
$\Q$&&$G=\Sp(1)\times\Sp(1)\times \textrm{O}(1,1)$&$G=\SU^\star(4)\times \Sp(1)$\\
&&$H=\Sp(1)\times\Sp(1)$&$H=\Sp(1) \times \Sp(2)$\\

&&&&\\

&&$\mathcal{N}=(2,1), f=128$&$\mathcal{N}=(2,2), f=256$\\
$\Oct$&&$G=\SU^\star(4)\times \Sp(1)$&$G=\SO(5,5)$\\
&&$H=\Sp(2) \times \Sp(1)$&$H=\Sp(2)\times\Sp(2)$\\
\\
\hline
\hline
\end{tabular}
\caption{\footnotesize{Second floor  of pyramid ($D=6$ supergravity).  The first row of each entry indicates the amount of supersymmetry $\mathcal{N}$ and the total number of degrees of freedom $f$. The second (third) row indicates the U-duality group $G$ (the maximal compact subgroup $H\subset G$) and its dimension. The scalar fields in each case parametrise the coset $G/H$.} }.
\label{tab:6Dsugra}
  \end{center}
\end{table}

\FloatBarrier

\paragraph{$D=10$ layer:} In $D=10$ we just have $\N=1$ SYM over $\Oct$, whose on-shell field content is a pair of octonions: a vector $\bold{8}_v$ and spinor $\bold{8}_s$ or $\bold{8}_c$ of $\SO(8)_{ST}$. When each Yang-Mills theory contains an $\bold{8}_s$ we apply the recipe as above:
\be\label{E8decomp}
\begin{split}
E_{8(8)}&\supset \SO(8)_{ST_L}\times\SO(8)_{ST_R}\\
\mathbf{248} &\to\big[\mathbf{(28,1)+(1,28)}+(\bold{8}_c,\bold{8}_c)\big]+\big[(\bold{8}_s,\bold{8}_s)+(\bold{8}_v,\bold{8}_v)\big],
\end{split}
\ee
where, once again, we use square brackets to divide the generators into those that belong to the maximal compact subgroup SO(16) and those that do not. We should again take the diagonal subgroup in $\SO(8)_{ST_L}\times\SO(8)_{ST_R}$, taking tensor products of the representations appearing in the two slots:
\be
\begin{split}
E_{8(8)}&\supset \SO(8)_{ST}\\
\mathbf{248} &\to\big[\mathbf{28+28+1}+\bold{28}_c+\bold{35}_c\big]+\big[\bold{1}+\bold{28}_s+\bold{35}_s+\bold{1}+\bold{28}_v+\bold{35}_v\big].
\end{split}
\ee
Discarding all but the spacetime singlets leaves us with a copy of $\SL(2,\R)$ decomposed into the trivial group,
\be
\bold{3}\rightarrow\big[\bold{1}]+[\bold{1}+\bold{1}\big],
\ee
so we recover the familiar $\SL(2,\R)$ U-duality of Type IIB supergravity.

To obtain Type IIA we just exchange $\bold{8}_s\leftrightarrow\bold{8}_c$ in the right-hand slots of \autoref{E8decomp}:
\be
\begin{split}
E_{8(8)}&\supset \SO(8)_{ST_L}\times\widetilde\SO(8)_{ST_R}\\
\mathbf{248} &\to\big[\mathbf{(28,1)+(1,28)}+(\bold{8}_c,\bold{8}_s)\big]+\big[(\bold{8}_s,\bold{8}_c)+(\bold{8}_v,\bold{8}_v)\big],
\end{split}
\ee
which becomes
\be
\begin{split}
E_{8(8)}&\supset \SO(8)_{ST}\\
\mathbf{248} &\to\big[\mathbf{28+28}+\bold{8}_v+\bold{56}_v\big]+\big[\bold{8}_v+\bold{56}_v+\bold{1}+\bold{28}_v+\bold{35}_v\big],
\end{split}
\ee
leaving a single non-compact $\bold{1}$ to generate O$(1,1)$. This is the correct U-duality, since there is only a single scalar in Type IIA, which lives on the scalar manifold $\R\cong\Orth(1,1)/\id$.

\begin{table}
 \begin{center}
\scriptsize
 \begin{tabular}{c|llllllllllll}
 \hline
\hline
 \\
$\alg_{8\N_L}\backslash\alg_{8\N_R}$&&$\Oct$\\
\\
\hline
\\
&&$\mathcal{N}=2$ (IIA), $f=256$\\
$\Oct$&&$G=\textrm{O}(1,1)$\\
&&$H=\mathds{1}$\\
\\
\hline
\hline
\end{tabular}
\quad\quad\quad\quad\quad
 \begin{tabular}{c|llllllllllll}
 \hline
\hline
 \\
$\alg_{8\N_L}\backslash\alg_{8\N_R}$&&$\Oct$\\
\\
\hline
\\
&&$\mathcal{N}=2$ (IIB), $f=256$\\
$\Oct$&&$G=\SL(2,\R)$\\
&&$H=\SO(2)$\\
\\
\hline
\hline
\end{tabular}
\caption{\footnotesize{Magic square of $D=10$ supergravity theories.  The left-hand (right-hand) table is obtained by tensoring  SYM of opposing (matching) chiralities, which is equivalent to applying a triality to the magic pyramid formula. Of course, there is no room for matter couplings in $D=10$.}}
\label{tab:10Dsugra}
  \end{center}
\end{table}

\FloatBarrier

\subsection{Complex and Quaternionic Structures}

\begin{figure}[t]
\begin{center}
\includegraphics[scale=0.12]{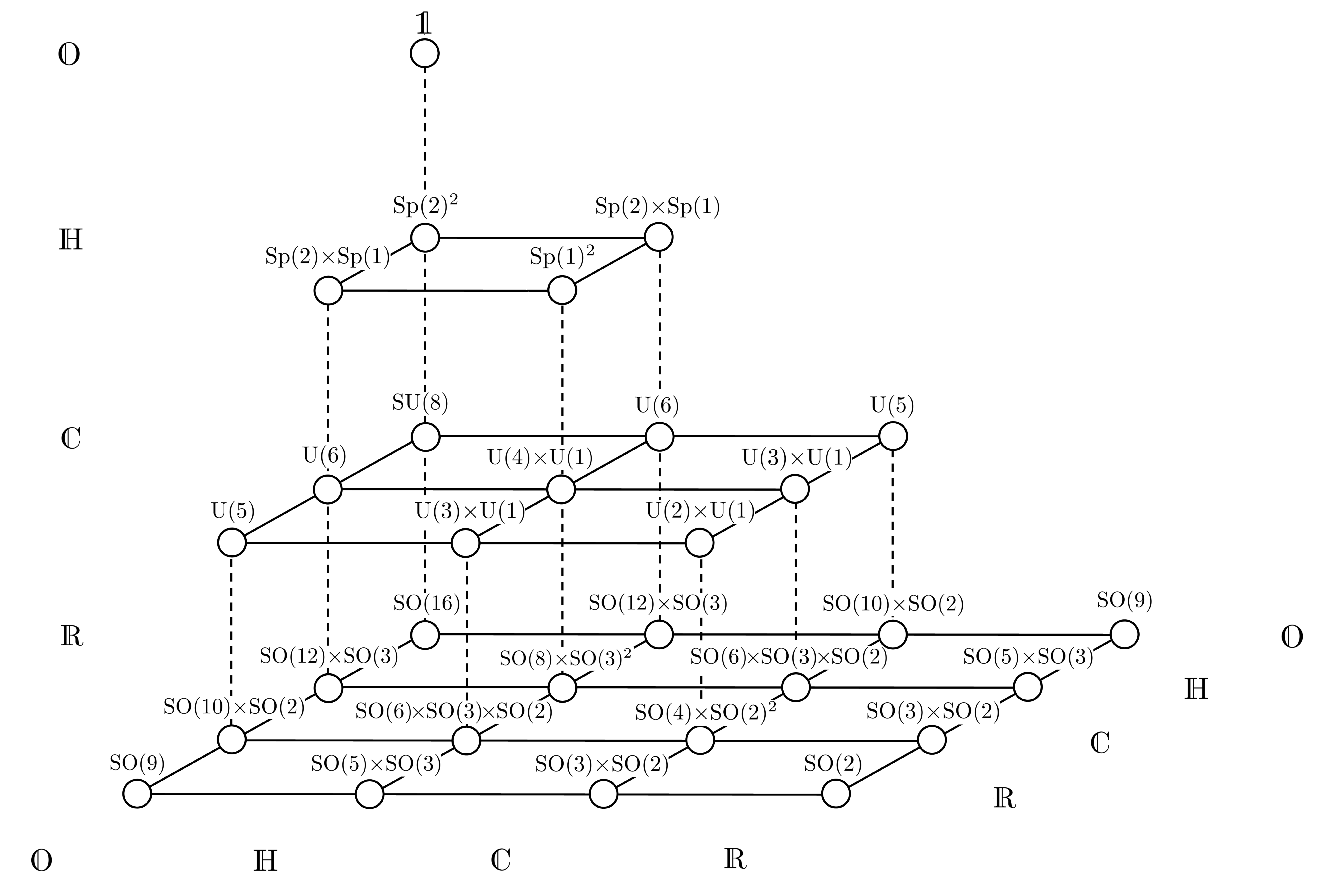}
\caption{{Magic pyramid of maximal compact subgroups}} \label{fig:Hp}
\end{center}
\end{figure}

It is interesting to look at the magic pyramid of maximal compact subgroups, shown in \autoref{fig:Hp}. The striking feature is that the $D=3$ square is built from orthogonal groups, the $D=4$ square from unitary groups and the $D=6$ square from symplectic groups. This is no mere coincidence:  $\SO(N)$ is the group of rotations in a real $N$-dimensional
space, $\Un(N)$ is the group of rotations in a complex $N$-dimensional space
and $\Sp(N)$ is the group of rotations in a quaternionic $N$-dimensional
space \cite{Baez:2001dm, Georgi:514148},
\be
\begin{split}
\mathfrak{so}(N)&=\{X\in\R[N]\hspace{0.1cm}|\hspace{0.1cm}X^\dagger=-X\},\\
\mathfrak{u}(N)&=\{X\in\C[N]\hspace{0.1cm}|\hspace{0.1cm}X^\dagger=-X\},\\
\mathfrak{sp}(N)&=\{X\in\Q[N]\hspace{0.1cm}|\hspace{0.1cm}X^\dagger=-X\},\\
\end{split}
\ee
where $\Al[N]$ denotes the set of $N\times N$ matrices with entries\footnote{Incidentally, this explains our insistence on referring to $\SU(2)$ as $\Sp(1)$, the group generated by $1\times 1$ anti-Hermitian quaternionic matrices.} in $\Al$. Note that (up to factors of SO(3) and SO(2)) as we climb the dotted lines of the pyramid, corresponding to dimensional oxidation, the maximal compact subgroups go as $\SO(\N_{D=3})=\SO(2\N_{D=4})\rightarrow\SU(\N_{D=4})$ when ascending from $D=3$ to $D=4$. These groups are of course the supergravity R-symmetries: $\SO(\N)$ in $D=3$ and $\SU(\N)$ in $D=4$. The R-symmetry groups are the automorphisms of the supersymmetry algebra, with the supercharges $Q$ transforming in the defining representation. Restricting the $D=3$ symmetries by demanding that they commute with the single generator $J$ of $\mf{u}(1)_{ST}$ amounts to demanding that the generators of $\SO(2\N_{D=4})$ commute with a complex structure, as $J$ satisfies $J^2Q=-\id Q$. From this point of view it is clear why we find $\SO(2\N_{D=4})\rightarrow\SU(\N_{D=4})$, since in general
\be
\mf{u}(N)\cong\left\{u\in\mf{so}(2N)\Big|[u,J]=0,~J^2=-\id,~J\in\mf{so}(2N)\right\},
\ee
where the U(1) factor of $\Un(N)\cong\SU(N)\times\Un(1)$ is generated by the complex structure $J$ itself. In our case the complex structure is the generator of the spacetime little group $\Un(1)_{ST}$.

To understand the different R-symmetry groups as we ascend from $D=3$ to $D=6$ we require the notion of a quaternionic structure. This is a triple of $4N\times 4N$ matrices $J_1$, $J_2$ and $J_3:=J_1J_2$ satisfying the quaternion algebra:
\be\label{QUATSTRUC}
J_iJ_j=-\delta_{ij}\id+\eps_{ijk}J_k,~~~i=1,2,3,
\ee
which we find belong to the Lie algebra $\mf{so}(4N)$. Just as $\Un(N)$ may be seen as the subgroup of $\SO(2N)$ that commutes with a complex structure, symplectic groups are the subgroups of $\SO(4N)$ that commute with the $J_i$:
\be
\mf{sp}(N)\cong\left\{u\in\mf{so}(4N)\Big|[u,J_1]=[u,J_2]=0,~J_1^2=J_2^2=-\id,~,J_1J_2=-J_2J_1,~J_1,J_2\in\mf{so}(4N)\right\}
\ee
(note that the conditions on $J_{1,2}$ in the curly brackets are enough to ensure that (\ref{QUATSTRUC}) is satisfied). Just as the complex structure $J$ generates $\mf{u}(1)$, the quaternionic structure matrices $J_i$ themselves generate a copy of $\mf{sp}(1)$, on account of (\ref{QUATSTRUC}), which by construction commutes with the $\mf{sp}(N)$. In $D=6$ the spacetime little group is $\SO(4)_{ST}\cong\Sp(1)^+_{ST}\times\Sp(1)^-_{ST}$, and we can understand each of these $\Sp(1)$ factors as being generated by a quaternionic structure. For example, in the maximal $(\N_+,\N_-)=(2,2)$ supergravity we have 16 supercharges, divided equally between the two chiralities, splitting the possible SO(16) group of transformations into $\SO(8)\times\SO(8)$. Putting a quaternionic structure on each of these SO(8) factors leaves us with an overall symmetry $\Sp(2)\times\Sp(2)\times\Sp(1)^+_{ST}\times\Sp(1)^-_{ST}$.

\subsection{S-duality of $\mathcal{N}=4$ SYM and supergravity}

When the tensor product involves at least one $\mathcal{N}=4$ SYM multiplet, it is tempting to speculate that the exact $\SL(2, \Z)$ S-duality of $\mathcal{N}=4$ SYM contributes to the S-duality of  the resulting supergravity. How this might actually work remains unclear, especially given the exchange of the gauge group for its GNO (Goddard, Nuyts, and Olive) dual\footnote{One possibility is that the left/right gauge groups must be GNO duals. We thank Neil Lambert for  sharing this suggestion.} \cite{Goddard:1976qe}. However, a minimal consistency requirement can be checked. The $\SL(2, \Z)_S$ S-duality of supergravity acts nontrivially on the NS-NS sector gauge potentials and their duals, which together transform as doublets. The RR sector potentials and their duals are, on the other hand, singlets. The NS-NS potentials are identified as those originating from $\phi_L\otimes A_{R\nu}$ and $A_{L\mu} \otimes \phi_R$ products, while the RR sector potentials come from spinor-spinor products $\lambda_L\otimes\lambda_R$ (consistent with the familiar type II  story in $D=10$). This  yields the following counting of NS-NS and RR potentials and  dual potentials,
\be
\begin{array}{l|lllll}
&[\mathcal{N}=4]_R &[\mathcal{N}=2]_R &[\mathcal{N}=1]_R &[\mathcal{N}=0]_R  \\[6pt]
\hline
&&&\\[-2pt]
[\mathcal{N}=4]_L &2\times 12+ 2\times 16 &2\times 8+ 2\times 8 &2\times 6+ 2\times 4 &2\times 6+ 0  \\[6pt]
[\mathcal{N}=2]_L & &2\times 4+ 2\times 4 &2\times 2+ 2\times 2 &2\times 2+ 0  \\[6pt]
[\mathcal{N}=1]_L & & &0+ 2\times 1 &0+ 0  \\[6pt]
[\mathcal{N}=0]_L & & & & 0 + 0  \\[6pt]
\end{array}
\ee

Decomposing the U-duality representations carried by the gauge potentials and their duals under the product of S and T dualities we have,
\be
\scriptsize
\begin{array}{c|llllllllll}
 &[\mathcal{N}=4]_R &[\mathcal{N}=2]_R &[\mathcal{N}=1]_R &[\mathcal{N}=0]_R  \\[6pt]
\hline
&&&\\[-2pt]
[\mathcal{N}=4]_L &\begin{array}{l}
E_{7(7)}\supset\\
\SL(2)\times \SO(6,6)\\
\rep{56}\rightarrow \\ \rep{(2, 12) + (1, 32)}\end{array} 
&\begin{array}{l}
\SO^\star(12)\supset\\
\SL(2)\times \SO(3)\times\SO(2,6)\\
\rep{32}\rightarrow \\  \rep{(2, 1, 8_{\text{s}}) + (1,  2, 8_{\text{c}})}\end{array} 
 &\begin{array}{l}
\SU(1,5)\supset\\
\SL(2)\times \SO(2)\times\SO(6)\\
\rep{20}\rightarrow \\ \rep{(2, 6)}_{0} + \rep{(1, 4)}_{3}+ \rep{(1, \bar{4})}_{-3}\end{array} 
 &\begin{array}{l}
\SL(2)\times \SO(6)\supset\\
\SL(2)\times \SO(6)\\
\rep{(2,6)}\rightarrow \\ \rep{(2, 6)}\end{array} 
\\[6pt]
&&&\\[-2pt]
[\mathcal{N}=2]_L & 
&\begin{array}{l}
\SL(2)\times \SO(6,2)\supset\\
 \SL(2)_S\times\SL(2)^2\times\SU(2)^2\\
\rep{(2,8)}\rightarrow \\ \rep{(2, 2, 2, 1, 1) + ( 1, 2, 1, 2, 2 )}\end{array} 
 &\begin{array}{l}
\Un(1, 3)\supset\\
\SL(2)\times\Un(1)\times \Un(2)\\
\rep{4}\rightarrow \\ \rep{(2, 1)_{1} + (1, 2)_{-1}}\end{array} 
 &\begin{array}{l}
 \SL(2)\times \SO(2)\supset\\
\SL(2)\times \SO(2)\\
\rep{2}_{1}+\rep{2}_{-1}\rightarrow \\\rep{2}_{1}+\rep{2}_{-1}\end{array} 
\\[6pt]
&&&\\[-2pt]
[\mathcal{N}=1]_L &
& &\begin{array}{l}
\Un(1, 2)\supset\\
\SL(2)\times\Un(1)\\
\rep{1}_{1}+\rep{1}_{-1}\rightarrow \\ \rep{1}_{1}+\rep{1}_{-1}\end{array} 
 &\begin{array}{l}
\SL(2)\supset\\
\SL(2)\\
-\end{array} 
\\[6pt]
&&&\\[-2pt]
[\mathcal{N}=0]_L&&
 &
 &\begin{array}{l}
\SL(2)\supset\\
\SL(2)\\
-\end{array} 
\\[6pt]
&&&\\[-2pt]
\end{array} 
\ee
demonstrating a splitting of the potentials into their NS-NS and RR sectors consistent with their  tensor origin.

Note,   the $\SL(2, \R)$ factor appearing in the   U-duality of $[\mathcal{N}=2]_L\otimes[\mathcal{N}=2]_R$, which yields $\mathcal{N}=4$ supergravity coupled to two vector multiplets,   is not, as one might naturally assume, the S-duality group since it mixes NS-NS and RR, as can be checked by regarding it as a consistent truncation of  $\mathcal{N}=8$ supergravity. However, the $\SL(2, \R)$ S-duality inside $E_{7(7)}$ is retained inside the $\SO(6,2)$ factor of the $\mathcal{N}=4$ theory since $\SL(2, \R)\times\SO(6,2)$ is not a subgroup of $\SO(6,6)$. Of course, the strong-weak dualities of $\mathcal{N}=2$ SYM theories are not exact\footnote{Unless they come coupled to extra matter multiplets. For example, the $\SU(2)$ $\mathcal{N}=2$ SYM coupled to four hypermultiplets transforming in the fundamental is believed to be exact \cite{Seiberg:1994rs}.} and, as such, their role in this context is even less clear.


\section{The conformal magic pyramid}\label{sec:conf}

Rather than uniformly  tensoring SYM in each dimension we may consider instead  the conformal theories: CSm in $D=3$, SYM in $D=4$ and tensor multiplets in $D=6$. It is not clear what the appropriate theory should be in $D=10$ and we leave this question for future work. 

The tensorings of CSm and SYM in $D=3$ yield the same results so it is only the tensor multiplets in $D=6$ that we need to treat here. As for left/right SYM,  composing tensor multiplets with opposing chiralities we  obtain pure supergravity,
\be
\begin{split}
[(2,0)_{tensor}^{L}]\times[(0,2)_{tensor}^{R}] &= [(2,2)_{sugra}],\\
[(2,0)_{tensor}^{L}]\times[(0,1)_{tensor}^{R}] &= [(2,1)_{sugra}],\\
[(1,0)_{tensor}^{L}]\times[(0,1)_{tensor}^{R}] &= [(1,1)_{sugra}],
\end{split}
\ee
reproducing \autoref{tab:6Dsugra}.

On the other hand, for left/right tensor multiplets  with matching chiralities we   obtain  the exotic non-gravitational SD-Weyl (self-dual-Weyl) multiplets coupled to tensor multipets, 
\be \label{eq:SDW}
\begin{split}
[(2,0)_{tensor}^{L}]\times[(2,0)_{tensor}^{R}] &= [(4,0)_{\text{\emph{SD-Weyl}}}]\\
[(2,0)_{tensor}^{L}]\times[(1,0)_{tensor}^{R}] &= [(3,0)_{\text{\emph{SD-Weyl}}}]\\
[(1,0)_{tensor}^{L}]\times[(1,0)_{tensor}^{R}] &= [(2,0)_{\text{\emph{SD-Weyl}}}]+[(2,0)_{tensor}].
\end{split}
\ee
The $[(2,0)_{tensor}^{L}]\times[(2,0)_{tensor}^{R}] = [(4,0)_{\text{\emph{SD-Weyl}}}]$ squaring is given explicitly in \autoref{tab:D6SDW}.  This theory is developed in some detail in \cite{Hull:2000zn, Chiodaroli:2011pp}. It is  non-gravitational  with highest spin field transforming as the $(\rep{5}, \rep{1})$ of the little group $\Sp(1)^+_{ST}\times\Sp(1)^-_{ST}$. The terminology ``SD-Weyl" derives from the fact that the $(\rep{5}, \rep{1})$ representation has the symmetry properties of a  four-dimensional Euclidean self-dual Weyl tensor when written with $\SO(4)$ indices, as described in \cite{Hull:2000zn}.

The scalars of the SD-Weyl multiplets appearing in \eqref{eq:SDW} parameterise the following cosets 
\be \label{eq:SDWsyms}
\begin{split}
[(4,0)_{\text{\emph{SD-Weyl}}}]\quad&\quad\frac{E_{6(6)}}{\Sp(4)},\\
[(3,0)_{\text{\emph{SD-Weyl}}}]\quad&\quad\frac{\SU^\star(6)}{\Sp(3)},\\
 [(2,0)_{\text{\emph{SD-Weyl}}}]+[(2,0)_{tensor}]\quad&\quad\frac{\Orth(5,1)}{\Sp(2)}.
\end{split}
\ee
Hence, by exchanging SYM multiplets with tensor multiplets the $D=6$ level of the pyramid is adjusted,
\be
\begin{split}
\frac{\SO(5,5)}{\Sp(2)\times\Sp(2)}&\longrightarrow \frac{E_{6(6)}}{\Sp(4)}\\
\frac{\SU^\star(4)\times\Sp(1)}{\Sp(2)\times\Sp(1)}&\longrightarrow \frac{\SU^\star(6)}{\Sp(3)}\\
\frac{\Sp(1)^2\times\Orth(1,1)}{\Sp(1)^2\times\Z_2}&\longrightarrow \frac{\Orth(5,1)}{\Sp(2)}
\end{split}
\ee
while the remaining levels are left unchanged. Interestingly, this has the consequence that the exterior faces of the pyramid, as presented in \autoref{fig:confp1}, are given by the original magic square cut across its diagonal.

\begin{figure}[t]
\begin{center}
\includegraphics[scale=0.12]{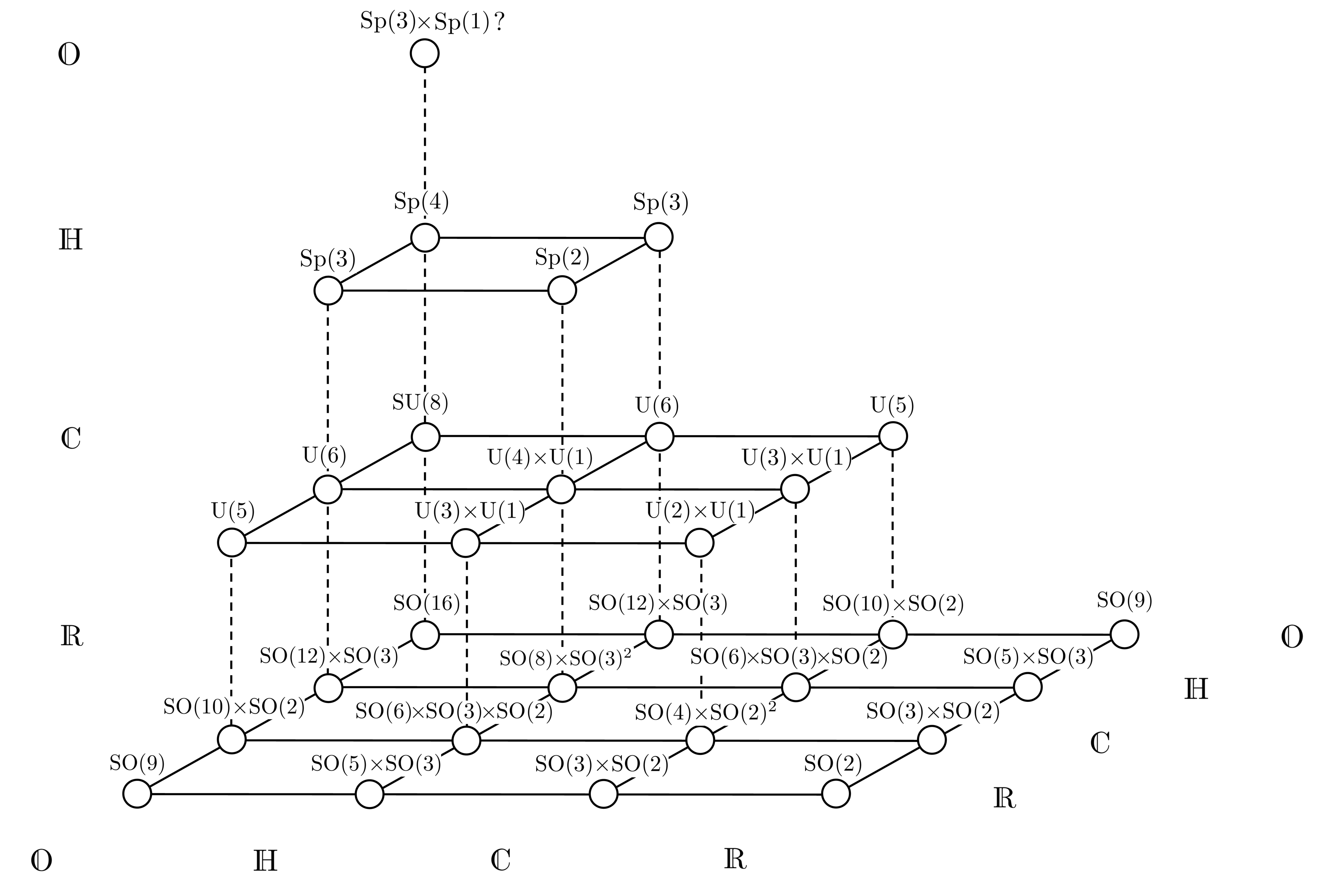}
\caption{{Conformal magic pyramid of maximal compact subgroups}}\label{CHpyr}
\end{center}
\end{figure}

An intriguing property of the tensor multiplets and SD-Weyl multiplets above is that every field is a scalar under $\Sp(1)^-_{ST}$, so the spacetime symmetry is essentially just $\Sp(1)^+_{ST}$ as long as we multiply tensor multiplets of a single chirality\footnote{Up until this point we have not mentioned $D=5$, but we note an interesting point about it here. Since the non-trivial little group in the $D=6$ chiral theories is $\Sp(1)^+_{ST}$, it becomes clear why the maximal $[(4,0)_{\text{\emph{SD-Weyl}}}]$ theory in $D=6$ and the maximal supergravity in $D=5$ both have have $E_{6(6)}$ as their U-duality groups: both may be obtained by restricting $E_{8(8)}$ to the subgroup that commutes with $SO(3)_{ST}\cong\Sp(1)^+_{ST}$.}. Mathematically the conformal pyramid is perhaps the most natural, since we can understand:
\begin{itemize}
\item the $D=4$ square as the Freudenthal-Rosenfeld-Tits magic square restricted to the subgroups that commute with a complex structure 
\item the $D=6$ square as the Freudenthal-Rosenfeld-Tits magic square restricted to the subgroups that commute with a single quaternionic structure (as opposed to the pair of quaternionic structures we found for the SYM-squared pyramid).
\end{itemize}
See \autoref{CHpyr}. From this perspective a method for obtaining the $(1,0)$ tensor multiplet
\be
B^+\in\text{Im}~\Q\sim\bold{(3,1)},~~~\phi\in\text{Re}~\Q\sim\bold{(1,1)},~~~\lambda\in\Q\sim\bold{(2,2)} ~~~\text{of}~~\Sp(1)^+_{ST}\times\Sp(1)
\ee
from the $(1,0)$ Yang-Mills multiplet 
\be
A\in\Q\sim\bold{(2,2,1)},~~~\lambda\in\Q\sim\bold{(2,1,2)}~~~\text{of}~~\Sp(1)^+_{ST}\times\Sp(1)^-_{ST}\times\Sp(1)
\ee
would be to identify $\Sp(1)^+_{ST}\sim\Sp(1)^-_{ST}$ and tensor product: $\bold{2}\times\bold{2}=\bold{3+1}$. So when $\Al_{4\N}=\Q$ the tensor multiplet is just that of Yang-Mills with positive and negative chiralities identified. The group $\Sp(1)^+$ just acts as orthogonal transformations on $\text{Im}~\Q$. For the $(2,0)$ tensor multiplet, $\Al_{4\N}=\Oct$, we have an $\Sp(1)^+_{ST}\times\Sp(2)$ overall symmetry (which simply comes from restricting SO(8) to the subgroup commuting with a quaternionic structure). This motivates the following definition of $\widetilde{\mf{sym}}$, our notation for the overall symmetry algebras of the conformal theories in $D=3,4,6$:
\be
\widetilde{\mf{sym}}(\Al_n,\Al_{n\N}):=\big\{(A,B,C)\in \mathfrak{tri}(\Al_{n\N})|~ [A, \mf{g}_{ST}]=0~~\forall A\notin \mf{g}_{ST}\big\},
\ee
where $a,b\in\Al_{n\N}$ and $\mf{g}_{ST}$ is the subalgbra of $\mathfrak{so}(\Al_{n\N})$ that acts as orthogonal transformations on $\Al_n\subset\Al_{n\N}$ when $n\neq 4$ and acts as othogonal transformations on $\text{Im}~\Al_n=\text{Im}~\Q\subset\Al_{4\N}$ when $n=4$. The fact that this definition is not democratic in the division algebras might seem unnatural, but the special treatment for $\Q$ just represents the additional requirement that the $D=6$ theories be completely chiral; the resulting algebras agree with $\mf{sym}$ in $D=3,4$ but $\widetilde{\mf{sym}}(\Q,\Q)\cong\mf{sp}(1)_{ST}\oplus\mf{sp}(1)$ and $\widetilde{\mf{sym}}(\Q,\Oct)\cong\mf{sp}(1)_{ST}\oplus\mf{sp}(2)$.

The U-dualities $\mf{u}$ of the conformal pyramid are then given by
\be\label{eq:conformalformula}
\mf{u}\cong\mathfrak{ConfPyr}(\alg_{n}, \alg_{n\susy_L}, \alg_{n\susy_R}):=\left\{u\in\mf{L}_3(\alg_{n\N_L}, \alg_{n\N_R})-\mf{g}_{ST}\Big|[u,\mf{g}_{ST}]=0\right\}.
\ee
In practice we find the groups of the conformal pyramid (in $D=3,4,6$) using the following method:
\begin{itemize}
\item Decompose $\mf{tri}(\Al_{n\N_{L}})\oplus\mf{tri}(\Al_{n\N_{R}})$ into $\widetilde{\mf{sym}}(\Al_n,\Al_{n\N_{L}})\oplus\widetilde{\mf{sym}}(\Al_n,\Al_{n\N_{R}})$,
\item Identify $\mf{g}_{ST}$ as  the diagonal subalgebra of $\mf{g}_{ST_L}\oplus\mf{g}_{ST_R}$,
\item Discard all generators that transform non-trivially under the spacetime symmetries $\mf{g}_{ST}$.
\end{itemize}
Once again, to find the maximal compact subgroups we just replace $\mf{L}_3(\alg_{n\N_L}, \alg_{n\N_R})$ with $\mf{L}_1(\alg_{n\N_L}, \alg_{n\N_R})$ in the above. While this method does not tell us how to obtain the U-duality in $D=10$,  we venture some speculations on this matter as part of our closing remarks in \autoref{sec:conclusion}.

\subsection{Barton-Sudbery-style formula}

For the compact subgroups $\mf{h}$ it is instructive to look at which generators in each of the three terms of $\mf{L}_1(\alg_{n\N_L}, \alg_{n\N_R}):=\mf{tri}(\alg_{n\N_L})\oplus \mf{tri}(\alg_{n\N_R})+(\alg_{n\N_L}\otimes\alg_{n\N_R})$ commute with $\mf{g}_{ST}$. For the first two terms,
\be
[\mf{tri}(\alg_{n\N_L})\oplus \mf{tri}(\alg_{n\N_R}),\mf{g}_{ST}]=0
\ee
is solved by 
\be
\mf{int}(\alg_{n\N_L})\oplus\delta_{2n}\mathfrak{u}(1)_{ST_{L}}\oplus \mf{int}(\alg_{n\N_R})\oplus\delta_{2n}\mathfrak{u}(1)_{ST_{R}},
\ee
where 
\be
\mf{int}(\Al_n,\Al_{n\N_{L/R}})=\widetilde{\mf{sym}}(\Al_n,\Al_{n\N_{L/R}})\ominus\mf{g}_{ST_{L/R}}.
\ee
The $\delta_{2n}\mathfrak{u}(1)_{ST_{L/R}}$ terms come from the fact that when $n=2$, $\mf{g}_{ST}\cong\mathfrak{u}(1)$ and so commutes with itself. The group we identify as spacetime in the supergravity theory is the diagonal subgroup of the left and right spacetime groups; subtracting this we are left with
\be
\mf{int}(\alg_{n\N_L})\oplus \mf{int}(\alg_{n\N_R})\oplus\delta_{2n}\mathfrak{u}(1).
\ee
Finally we denote the solution to
\be
[\alg_{n\N_L}\otimes\alg_{n\N_R},\mf{g}_{ST}]=0
\ee
(slightly schematically) by 
\be
\Al_n(\Al_{\N_L}\ox\Al_{\N_R})
\ee
since its dimension is $n\N_L\N_R$, and this notation captures its essence; we have made $\Al_{n\N_{L/R}}$ look like $\sim\Al_n\Al_{\N_{L/R}}$, and then brought the left and right pieces together, identifying a diagonal $\Al_n$ algebra. Putting all of this together we arrive at a Barton-Sudbery-style formula for the compact subgroups of the conformal pyramid
\be
\mathfrak{h}=\mf{int}(\Al_n,\Al_{n\N_L})\oplus\mf{int}(\Al_n,\Al_{n\N_R})+\Al_n(\Al_{\N_L}\ox\Al_{\N_R})+\delta_{2n}\mathfrak{u}(1),
\ee
which allows one to build up the symmetries of the squared theories from those of the left/right conformal theories.

\section{Conclusions}\label{sec:conclusion}

We began with the observation developed in \cite{Anastasiou:2013cya} that   $\mathcal{N}=2^m$-extended  SYM theories in $D=n+2$ spacetime dimensions may be formulated with a single Lagrangian and single set of transformation rules, but with spacetime fields valued in $\alg_{n\mathcal{N}}$. This perspective reveals a role for the   triality algebras; once the fields are regarded as division algebras consistency  with supersymmetry constrains the possible space of transformations to $\mathfrak{sym}(\alg_n, \alg_{n\mathcal{N}})\subseteq \mathfrak{tri}(\alg_{n\mathcal{N}})$. 

Tensoring left/right SYM multiplets valued in $\alg_{n\mathcal{N}_{L/R}}$ then naturally leads us to $\mathcal{N}_L+\mathcal{N}_R$ supergravity multiplets with spacetime fields valued in $\alg_{n\mathcal{N}_{L}}\otimes\alg_{n\mathcal{N}_{R}}$. For $D=1+2$ this yields  a set of supergravities with U-duality groups given by the magic square of Freudenthal-Rosenfeld-Tits. For $n=2,4,8$, identifying a common spacetime subalgebra $\alg_n$ truncates the magic square to a $3\times 3$, $2\times 2$, and $1\times 1$ array of subalgebras,  corresponding precisely to the  U-dualities obtained by tensoring SYM  in $D=4,6$ and $10$, respectively. Together the four ascending squares constitute a magic pyramid of algebras defined by the magic pyramid formula \eqref{eq:pyramid}.  The exceptional octonionic row and column of each level is constrained by supersymmetry to give the unique  supergravity multiplet. On the other hand, the interior $3\times 3$, $2\times2$, $1\times1$ and $0\times0$ squares can and do admit matter couplings.  These additional matter  multiplets are just as required to give the U-dualities predicted by the pyramid formula. Interestingly, in these cases the degrees of freedom are split evenly between the graviton multiplet and the matter multiplets, the number  of which  is determined by the rule\footnote{We thank Andrew Thomson for pointing out this rule. Note  the subtlety  in $D=6$ that one must treat $\N^+$ and $\N^-$ separately. Hence, for example, $[(1,0)]\times [(0,1)]$ has $k=0$.} $k=\min(\N_L, \N_R)$.

The magic pyramid supergravity theories are rather non-generic. Not only are they, in a sense,   defined by the magic pyramid formula, they are also generated by tensoring the division algebraic SYM multiplets.   It would therefore be interesting to explore whether they collectively possess other special properties, particularly    as quantum theories, which can be traced back to their magic square origins.  For example, in the maximal $[\mathcal{N}_{L}=4\;{\text{SYM}}] \times[\mathcal{N}_{R}=4\;{\text{SYM}}]$ case it has  been shown that $\mathcal{N}=8$ supergravity is four-loop finite \cite{Bern:2009kd}, a result which cannot be attributed to supersymmetry alone. While $\mathcal{N}=8$ is expected to have the best possible UV behaviour, as suggested by its  connection to $\mathcal{N}=4$ SYM, it could still be that the remaining magic square supergravities  share some structural features due to their common gauge $\times$ gauge origin and closely related global symmetries.

Conversely, one might also seek  extensions of the magic pyramid  construction which could account for more generic  supergravities.  The magic supergravities of Gunaydin, Sierre and Townsend \cite{Gunaydin:1983rk, Gunaydin:1983bi}, for example, admit at least one obvious generalisation using the  family of spin-factor Jordan algebras, suggesting a possible extension of the present construction by incorporating matter multiplets.

Let us return  to the present treatment, now in the conformal case.  In \autoref{sec:conf} we saw that  tensoring the conformal theories in $D=3,4,6$ resulted in a pyramid with the intriguing feature that its exterior faces are given by the Freudenthal-Rosenfeld-Tits square cut across its diagonal. In particular, ascending up the maximal spine one encounters  the famous exceptional sequence $E_{8(8)}, E_{7(7)}, E_{6(6)}$, but where $E_{6(6)}$ belongs to the exotic $(4, 0)$ theory in $D=6$.  This pattern   suggests the existence of some highly exotic $D=10$ theory with $F_{4(4)}$ U-duality group\footnote{Note, $F_{4(4)}/\Sp(3)\times\Sp(1)$ also appears in three dimensions as the scalar coset of the $\N=4$ magic supergravity coupled to six vector multiplets.  It corresponds to  dimensional reduction of the $D=4, \N=2$ magic supergravity based on the Jordan algebra of $3\times 3$ real Hermitian matrices \cite{Gunaydin:1983rk}. We thank one of the referees for bringing this observation to our attention.}. We would naturally require it to dimensionally reduce to the $(4, 0)$ theory in $D=6$ on some non-trivial manifold (orbifold), consistent with scalars living in $E_{6(6)}/\Sp(4)$ and $F_{4(4)}/\Sp(3)\times\Sp(1)$ in $D=6$ and $10$, respectively. The $D=6$ supercharges transform as the $\rep{8}$ of $\Sp(4)$, which breaks to the $\rep{(6,1)+(1, 2)}$ of $\Sp(3)\times\Sp(1)$, leaving $\mathcal{N}=2,6,8$ as possibilities in $D=10$. Naively at least,  $\mathcal{N}=2$  is ruled out by the standard classification of supermultiplets \cite{Nahm:1977tg, Strathdee:1986jr} due to  the assumption that it may be dimensionally reduced to $D=6$, $\mathcal{N}=(4,0)$, since this would imply fields of helicity greater than 2  when dimensionally reducing on a 6-torus. If,  however, the $F_{4(4)}$ theory had some exotic dynamics  which broke the usual spacetime little group to some subgroup this logic may not hold. Taking into account the desired $F_{4(4)}$, this line of reasoning suggests a $G_2$ little group as one possibility.   These avenues will be explored elsewhere. 

There is, however, an obvious alternative interpretation of the conformal pyramid including its $F_{4(4)}$ tip. The $(4, 0)$ theory in $D=6$ with $E_{6(6)}$ U-duality reduces on a circle to $D=5$, $\mathcal{N}=8$ supergravity, again with $E_{6(6)}$ U-duality. The same result holds for the remaining three slots of the $D=6$  square; they each reduce to a $D=5$ supergravity theory with very same U-duality group. This is a consequence of the fact that the $D=6$ fields are singlets under the second factor of the little group $\Sp(1)\times\Sp(1)$, which therefore effectively reduces to the $D=5$ little group $\Sp(1)$. Each of the resulting $D=5$ theories may be obtained by squaring. Hence, \autoref{fig:confp1} may be regarded as a squashed  pyramid of  U-dualities for theories in $D=3,4,5,6$. The apex is now given by the  $D=6, \mathcal{N}=(3,1)$ theory, obtained from $[(2,0)^{L}_{tensor}]\times [(1,1)^{R}_{vector}]$, with $G/H$ given by $F_{4(4)}/\Sp(3)\times\Sp(1)$, as expected. Note, however, this multiplet contains gravitini but no graviton and is therefore not expected to define a consistent interacting theory.

We conclude with some brief remarks on the geometrical interpretation of the magic pyramid. When we made the observation that the U-dualities of the magic pyramid could be regarded as the isometries of the Lorentzian projective planes $(\alg_{\mathcal{N}_L}\otimes\alg_{\mathcal{N}_R})\mathds{P}^2$ (or submanifolds thereof), this was meant rather loosely in the cases of $\Q\otimes\Oct$ and $\Oct\otimes \Oct$, as they do not obey the axioms  of projective geometry. Unlike $\R\otimes \Oct$, $\Q\otimes\Oct$ and $\Oct\otimes \Oct$ are not division preventing a direct projective construction  and (unlike $\C\otimes\Oct$) Hermitian $3\times 3$ matrices over $\Q\otimes\Oct$ or $\Oct\otimes \Oct$ do not form a simple Jordan algebra, so the usual identification of points  (lines) with trace 1 (2) projection operators cannot be made \cite{Baez:2001dm}. Nonetheless, they are in fact geometric spaces, generalising projective spaces,    known as ``buildings'', on which  the U-dualities act as isometries. Buildings where originally introduced by Jacques Tits to  provide a geometric  approach to simple Lie groups, in particular the exceptional cases, but have since had far reaching implications. See, for example, \cite{Tits:1974, Buildings:2002} and the references therein. Of course, it has long been  known that increasing supersymmetry restricts the spaces on which the scalar fields may live,  as comprehensively demonstrated for $D=3$ in \cite{deWit:1992up}. Here we see that these restrictions lead us to the concept of buildings.  It may be of interest to examine whether this relationship between supersymmetry and buildings has some useful implications.

\FloatBarrier

\section*{Acknowledgments}

We would like to thank Bianca Cerchiai, Sergio Ferrara and  Alessio Marrani for useful discussions. We thank John Carrasco and Lance Dixon for discussions concerning the role of S-duality. We thank Zvi Bern for his encouragement. The work of LB is supported by an Imperial College Junior Research Fellowship. The work of MJD is supported by the STFC under rolling grant ST/G000743/1.  

\appendix

\section{$D=6$ Tensoring tables}\label{sec:6dtensor}
\FloatBarrier
In \autoref{tab:D6N8}, \autoref{tab:D6N6} and \autoref{tab:D6N4} we perform the $D=6$ SYM squaring on-shell to arrive at the supergravity and matter content. In each table the  fields are shown together with their little group $\SO(4)\cong\Sp(1)_+\times\Sp(1)_-$ representations. Note, we restrict to the semi-simple part of the full little group for massless states. The $[(2,0)_{tensor}^{L}]\times[(2,0)_{tensor}^{R}] = [(4,0)_{\text{\emph{SD-Weyl}}}]$ tensoring is given as an example in \autoref{tab:D6SDW}. The little group representations appearing in the left-handed SD-Weyl multiplets are given by $(\rep{n}, \rep{1})$, where $\rep{n}=\rep{1}, \rep{2},\ldots\rep{5}$. These irreps are carried by totally symmetric rank $\rep{n}-1$ tensors of $\Sp(1)_+$:
\be
(\rep{3}, \rep{1})\; B^{+}_{(A_1A_2)};\quad (\rep{4}, \rep{1})\; C^{+}_{(A_1A_2A_3)};\quad(\rep{5}, \rep{1})\; D^{+}_{(A_1A_2A_3A_4)}.
\ee 
The multiplicities are given by the dimension of the R-symmetry representation of the fields.

Consulting \autoref{tab:D6SDW} we see that there are 27 self-dual two-form field strengths transforming as the fundamental $\rep{27}$ of $E_{6(6)}$. There are 42 scalars parametrising $E_{6(6)}/\Sp(4)$. The fermonic fields, $C^+$ and $\lambda^+$, transform as the $\rep{8}$ and $\rep{48}$ of $\Sp(4)$ respectively.

\begin{table}[h]
\small
\begin{center}
 $\begin{array}{c|c|c|c|c}
 &\begin{array}{c}  {A}_\mu \\(\rep{2},\rep{2})\end{array}& \begin{array}{c}2  \lambda^{+} \\ 2(\rep{2},\rep{1}) \end{array}&\begin{array}{c}2 \lambda^{-} \\ 2(\rep{1},\rep{2})\end{array}&\begin{array}{c}4 {\phi} \\ 4(\rep{1},\rep{1})\end{array}\\
\hline
&&&&\\
\begin{array}{c}{A}_\mu\\ (\rep{2},\rep{2})\end{array}
&\begin{array}{cccccc} g_{\mu\nu} &+& B^{+}_{\mu\nu}+B^{-}_{\mu\nu}&+&\varphi \\ (\rep{3},\rep{3}) &+& (\rep{3},\rep{1})+(\rep{1},\rep{3})&+&(\rep{1},\rep{1})\end{array} 
&\begin{array}{cccccc} 2[\Psi_{\mu}^{-} &+& \chi^{-}]\\ 2[(\rep{3},\rep{2}) &+& (\rep{1},\rep{2})]\end{array} 
&\begin{array}{cccccc} 2[\Psi_{\mu}^{+} &+& \chi^{+}]\\ 2[(\rep{2},\rep{3}) &+& (\rep{2},\rep{1})]\end{array} 
&\begin{array}{cccccc} 4A_\mu \\ 4(\rep{2},\rep{2})\end{array} 
\\

&&&&\\
\begin{array}{c}2 {\lambda}^{+} \\ 2(\rep{2},\rep{1}) \end{array}
&\begin{array}{cccccc} 2[\Psi_{\mu}^{-} &+& \chi^{-}]\\ 2[(\rep{3},\rep{2}) &+& (\rep{1},\rep{2})]\end{array} 
&\begin{array}{cccccc} 4[\varphi &+& B_{\mu\nu}^{+}]\\ 4[(\rep{1},\rep{1}) &+& (\rep{3},\rep{1})]\end{array} 
&\begin{array}{cccccc} 4A_\mu \\ 4(\rep{2},\rep{2})\end{array} 
&\begin{array}{cccccc} 8\chi^{+} \\ 8(\rep{2},\rep{1})\end{array} 

\\

&&&&\\

\begin{array}{c}2 {\lambda}^{-} \\ 2(\rep{1},\rep{2})\end{array}
&\begin{array}{cccccc} 2[\Psi_{\mu}^{+} &+& \chi^{+}]\\ 2[(\rep{2},\rep{3}) &+& (\rep{2},\rep{1})]\end{array} 
&\begin{array}{cccccc} 4A_\mu \\ 4(\rep{2},\rep{2})\end{array} 
&\begin{array}{cccccc} 4[\varphi &+& B_{\mu\nu}^{-}]\\ 4[(\rep{1},\rep{1}) &+& (\rep{1},\rep{3})]\end{array} 
&\begin{array}{cccccc} 8\chi^{-} \\ 8(\rep{1},\rep{2})\end{array} 

\\

&&&&\\

\begin{array}{c}4{\phi} \\4 (\rep{1},\rep{1})\end{array}
&\begin{array}{cccccc} 4A_\mu \\ 4(\rep{2},\rep{2})\end{array} 
&\begin{array}{cccccc} 8\chi^{+} \\ 8(\rep{2},\rep{1})\end{array} 
&\begin{array}{cccccc} 8\chi^{-} \\ 8(\rep{1},\rep{2})\end{array} 
&\begin{array}{cccccc} 16\varphi \\ 16(\rep{1},\rep{1})\end{array} 

\\

\end{array}$
\caption{$D=6$, $[(1,1)_{SYM}^{L}]\times[(1,1)_{SYM}^{R}] = [(2,2)_{sugra}]$.}\label{tab:D6N8}
\end{center}
\end{table}

\begin{table}[h]
\begin{center}
 $\begin{array}{c|c|c}
 &\begin{array}{c} {A}_\mu \\(\rep{2},\rep{2})\end{array}& \begin{array}{c}2  \lambda^{-} \\ 2(\rep{1},\rep{2}) \end{array}\\
\hline
&&\\
\begin{array}{c}{A}_\mu\\ (\rep{2},\rep{2})\end{array}
&\begin{array}{cccccc} g_{\mu\nu} &+& B^{+}_{\mu\nu}+B^{-}_{\mu\nu}&+&\varphi \\ (\rep{3},\rep{3}) &+& (\rep{3},\rep{1})+(\rep{1},\rep{3})&+&(\rep{1},\rep{1})\end{array} 
&\begin{array}{cccccc} 2[\Psi_{\mu}^{+} &+& \chi^{+}]\\ 2[(\rep{2},\rep{3}) &+& (\rep{2},\rep{1})]\end{array} 
\\

&&\\
\begin{array}{c}2\lambda^{+} \\ 2(\rep{2},\rep{1}) \end{array}
&\begin{array}{cccccc} 2[\Psi_{\mu}^{-} &+& \chi^{-}]\\ 2[(\rep{3},\rep{2}) &+& (\rep{1},\rep{2})]\end{array} 
&\begin{array}{cccccc} 4A_\mu \\ 4(\rep{2},\rep{2})\end{array} 

\\

&&\\

\begin{array}{c}2\lambda^{-} \\ 2(\rep{1},\rep{2})\end{array}
&\begin{array}{cccccc} 2[\Psi_{\mu}^{+} &+& \chi^{+}]\\ 2[(\rep{2},\rep{3}) &+& (\rep{2},\rep{1})]\end{array} 
&\begin{array}{cccccc} 4[\varphi &+& B_{\mu\nu}^{-}]\\ 4[(\rep{1},\rep{1}) &+& (\rep{1},\rep{3})]\end{array} 
\\

&&\\

\begin{array}{c}4{\phi} \\4 (\rep{1},\rep{1})\end{array}
&\begin{array}{cccccc} 4A_\mu \\ 4(\rep{2},\rep{2})\end{array} 
&\begin{array}{cccccc} 8\chi^{-} \\ 8(\rep{1},\rep{2})\end{array} 

\\

\end{array}$
\caption{$D=6$, $[(1,1)_{SYM}^{L}]\times[(1,0)_{SYM}^{R}] = [(2,1)_{sugra}]$ }\label{tab:D6N6}
\end{center}
\end{table}

\begin{table}
\begin{center}
 $\begin{array}{c|c|c}
&\begin{array}{c} {A}_\mu \\(\rep{2},\rep{2})\end{array}& \begin{array}{c}2  \lambda^{-} \\ 2(\rep{1},\rep{2}) \end{array}\\
\hline
&&\\
\begin{array}{c}{A}_\mu\\ (\rep{2},\rep{2})\end{array}
&\begin{array}{cccccc} g_{\mu\nu} &+& B^{+}_{\mu\nu}+B^{-}_{\mu\nu}&+&\varphi \\ (\rep{3},\rep{3}) &+& (\rep{3},\rep{1})+(\rep{1},\rep{3})&+&(\rep{1},\rep{1})\end{array} 
&\begin{array}{cccccc} 2[\Psi_{\mu}^{+} &+& \chi^{+}]\\ 2[(\rep{2},\rep{3}) &+& (\rep{2},\rep{1})]\end{array} 
\\

&&\\

\begin{array}{c}2\lambda^{-} \\ 2(\rep{1},\rep{2})\end{array}
&\begin{array}{cccccc} 2[\Psi_{\mu}^{+} &+& \chi^{+}]\\ 2[(\rep{2},\rep{3}) &+& (\rep{2},\rep{1})]\end{array} 
&\begin{array}{cccccc} 4[\varphi &+& B_{\mu\nu}^{-}]\\ 4[(\rep{1},\rep{1}) &+& (\rep{1},\rep{3})]\end{array} 
\\

\end{array}$
\caption{$D=6$, $[(1,0)_{SYM}^{L}]\times[(1,0)_{SYM}^{R}] = [(2,0)_{sugra}]+[(2,0)_{tensor}]$ }\label{tab:D6N4}
\end{center}
\end{table}
\FloatBarrier

\begin{table}[h]
\begin{center}
 $\begin{array}{c|c|c}
&\begin{array}{c} {A}_\mu \\(\rep{2},\rep{2})\end{array}& \begin{array}{c}2  \lambda^{+} \\ 2(\rep{2},\rep{1}) \end{array}\\
\hline
&&\\
\begin{array}{c}{A}_\mu\\ (\rep{2},\rep{2})\end{array}
&\begin{array}{cccccc} g_{\mu\nu} &+& B^{+}_{\mu\nu}+B^{-}_{\mu\nu}&+&\varphi \\ (\rep{3},\rep{3}) &+& (\rep{3},\rep{1})+(\rep{1},\rep{3})&+&(\rep{1},\rep{1})\end{array} 
&\begin{array}{cccccc} 2[\Psi_{\mu}^{-} &+& \chi^{-}]\\ 2[(\rep{3},\rep{2}) &+& (\rep{1},\rep{2})]\end{array} 
\\

&&\\

\begin{array}{c}2\lambda^{-} \\ 2(\rep{1},\rep{2})\end{array}
&\begin{array}{cccccc} 2[\Psi_{\mu}^{+} &+& \chi^{+}]\\ 2[(\rep{2},\rep{3}) &+& (\rep{2},\rep{1})]\end{array} 
&\begin{array}{cccccc} 4[A_\mu]\\ 4[(\rep{2},\rep{2})]\end{array} 
\\

\end{array}$
\caption{$D=6$, $[(1,0)_{SYM}^{L}]\times[(0, 1)_{SYM}^{R}] = [(1,1)_{sugra}]$ }\label{tab:D6N4}

\end{center}
\end{table}

\begin{table}[h]
\begin{center}

 $\begin{array}{c|c|c|c}
&\begin{array}{c}  {B}^+ \\(\rep{3},\rep{1})\end{array}& \begin{array}{c}4  \lambda^{+} \\ 4(\rep{2},\rep{1}) \end{array}&\begin{array}{c} 5 \phi \\ (\rep{1},\rep{1})\end{array}\\
\hline
&&&\\
\begin{array}{c}{B^+}\\ (\rep{3},\rep{1})\end{array}
&\begin{array}{cccccc} D^+ &+& B^{+}&+&\varphi \\ (\rep{5},\rep{1}) &+& (\rep{3},\rep{1})&+&(\rep{1},\rep{1})\end{array} 
&\begin{array}{cccccc} 4[C^{+} &+& \chi^{+}]\\ 4[(\rep{4},\rep{1}) &+& (\rep{2},\rep{1})]\end{array} 
&\begin{array}{cccccc} 5 B^+\\ 5(\rep{3},\rep{1}) \end{array} 
\\

&&&\\
\begin{array}{c}4 {\lambda}^{+} \\ 4(\rep{2},\rep{1}) \end{array}
&\begin{array}{cccccc} 4[C^{+} &+& \chi^{+}]\\ 4[(\rep{4},\rep{1}) &+& (\rep{2},\rep{1})]\end{array} 
&\begin{array}{cccccc} 16[B^+ &+& \varphi]\\ 16[(\rep{3},\rep{1}) &+& (\rep{1},\rep{1})]\end{array} 
&\begin{array}{cccccc} 20 \chi^{+} \\ 20(\rep{2},\rep{1})\end{array} 

\\

&&&\\

\begin{array}{c}5 \phi \\ 5(\rep{1},\rep{1})\end{array}
&\begin{array}{cccccc} 5 B^+\\ 5(\rep{3},\rep{1}) \end{array} 
&\begin{array}{cccccc} 20 \chi^{+} \\ 20(\rep{2},\rep{1})\end{array} 
&\begin{array}{cccccc} 25\varphi \\ (\rep{1},\rep{1})\end{array}

\end{array}$
\caption{$D=6$, $[(2,0)_{tensor}^{L}]\times[(2,0)_{tensor}^{R}] = [(4,0)_{\text{\emph{SD-Weyl}}}]$.}\label{tab:D6SDW}
\end{center}

\end{table}

\FloatBarrier


\providecommand{\href}[2]{#2}\begingroup\raggedright\endgroup

\end{document}